\documentclass[aps,pra,twocolumn,showpacs,superscriptaddress,floatfix]{revtex4-1}
\usepackage[utf8]{inputenc} 
\usepackage{amssymb,amsmath, nicefrac}
\usepackage{graphicx}
\usepackage[ruled]{algorithm2e}
\usepackage{ragged2e} 
\usepackage{braket}
\usepackage{siunitx}
\usepackage{xcolor}
\usepackage[normalem]{ulem}
\usepackage{bm}

\def\Rb87{^{87}\mathrm{Rb}}                     

\def\ex{\mathbf{e}_x}  
\def\ey{\mathbf{e}_y}  
\def\ez{\mathbf{e}_z}  
\newcommand{\e}[1]{\ensuremath{\times 10^{#1}}}

\begin{document}

\title{Self-Bayesian Aberration Removal via Constraints for Ultracold Atom Microscopy}

\author{Emine~Altunta\c{s}}
\email{altuntas@umd.edu}
\affiliation{Joint Quantum Institute, National Institute of Standards and Technology, and University of Maryland, Gaithersburg, Maryland, 20899, USA}
\author{I.~B.~Spielman}
\email{ian.spielman@nist.gov}
\affiliation{Joint Quantum Institute, National Institute of Standards and Technology, and University of Maryland, Gaithersburg, Maryland, 20899, USA}
\homepage{http://ultracold.jqi.umd.edu}
\date{\today}

\begin{abstract}
High-resolution imaging of ultracold atoms typically requires custom high numerical aperture (NA) optics, as is the case for quantum gas microscopy.
These high NA objectives involve many optical elements each of which contributes to loss and light scattering, making them unsuitable for quantum back-action limited ``weak'' measurements.
We employ a low cost high NA aspheric lens as an objective for a practical and economical---although aberrated---high resolution microscope to image $\Rb87$ Bose-Einstein condensates.
Here, we present a novel methodology for digitally eliminating the resulting aberrations that is applicable to a wide range of imaging strategies and requires no additional hardware.
We recover nearly the full NA of our objective, thereby demonstrating a simple and powerful digital aberration correction method for achieving optimal microscopy of quantum objects.
This reconstruction relies on a high quality measure of our imaging system's even-order aberrations from density-density correlations measured with differing degrees of defocus.
We demonstrate our aberration compensation technique using phase contrast imaging, a dispersive imaging technique directly applicable to quantum back-action limited measurements.
Furthermore, we show that our digital correction technique reduces the contribution of photon shot noise to density-density correlation measurements which would otherwise contaminate the desired quantum projection noise signal in weak measurements.
\end{abstract}

\maketitle


In many fields of study---from biophysics~\cite{Zhang2018} and medicine to astrophysics~\cite{Molina2001, Starck2002} and atomic physics~\cite{Ketterle96}---images are a key source of data, making high quality imaging systems essential. 
In all cases experimenters desire the maximum possible information from their images: imaging apertures limit the detected information; system inefficiencies discard information; and aberrations obfuscate what is finally detected.
In optics, sophisticated multi-element (and high cost) objectives are able to image objects with resolutions approaching fundamental limits~\cite{Zhang2019}.
In many cases, either because of technical incompatibilities, conflicting requirements or simply expense, these objectives cannot be employed.
In quantum gas experiments the object is an atomic sample encased in an ultra-high vacuum system that introduces aberrations and limits optical access.
Here we describe a versatile microscope for cold-atom imaging that fully uses the available optical access with low-cost optical elements in conjunction with a novel image reconstruction method, giving a combined hardware/software system that recovers near-diffraction limited performance.

Even ``quantum gas microscopes''~\cite{Bakr2009,Sherson2010}, the highest resolution imaging systems employed in cold-atom experiments, use algorithmic reconstruction techniques. 
These systems employ custom designed, high numerical aperture (NA) objectives to detect individual atoms in optical lattices by detecting their incoherent fluorescence.
The distribution of atoms can be reconstructed using algorithms similar to the CLEAN algorithm~\cite{Hogbom1974} from radio astronomy that construct distributions of point sources that are most consistent with the data given the system's point spread function (PSF).

In the case of coherent imaging, the observed aberrated images of cold atoms are related to the desired aberration-free images by multiplication of a contrast transfer function (CTF) in the spectral (Fourier) domain.
Because the CTF can reduce or eliminate the signal at some wavevectors, information is lost and direct inversion is not possible.
This can be resolved with a pseudo-inverse that uses a Bayesian prior in the vicinity of wavevectors with large information loss~\cite{Idier08, Demoment89}, but the resulting reconstructions suffer from artifacts and added noise~\cite{Turner2005,Wigley2016a,Perry2021}.

Inspired by the application of constraints to the phase retrieval problem in optics~\cite{Fienup1982, Fienup1993}, we present a new and versatile method that reduces artifacts in reconstructed images, while increasing the signal-to-noise ratio (SNR).
Spatially compact systems have a finite spectral width.
In the vicinity of the zeros in the CTF, our method effectively uses as a prior the weighted average of data from nearby wavevectors that are unresolvable given the system's assumed spectral width.
As such we have introduced a finite size constraint to the problem of refocusing and correcting aberrations in images of ultracold atoms.

In many cold-atom experiments further information is contained in density fluctuations often parameterized by the power spectral density (PSD). 
Examples sources of correlations include thermal noise at finite-temperature, quantum fluctuations at zero temperature, or quantum projection noise from the measurement process.
Other noise sources parasitically contribute to the PSD in experiments, and in our case photon shot noise is the largest such contributor.
In perfect imaging systems, this can be minimized by first windowing the data to contain only the region with atoms; however, in defocused or highly aberrated systems the atom signal is dispersed over much of the sensor and windowing becomes impractical.
We show that our aberration correction method overcomes this: by correcting for aberrations we first recover near-perfect images that then can be windowed to minimize the contribution of photon shot noise.

Our data consists of images of ultracold atom ensembles of roughly $10^5$ atoms that both phase-shift and absorb an illuminating probe beam.
Together the absorption and phase shift encode the density of atoms integrated along the propagation direction of the probe beam giving a 2D image of atomic density that we denote as an abstract ``data'' vector ${\bf d}$.
We focus on linear imaging systems where, as we describe below, aberrations and losses can be encoded as a linear transform described by the operator ${\bf H}$, the CTF.
Here the actual measurement outcome ${\bf m}$ is related to the desired data via the linear transformation ${\bf m} = {\bf H}\ {\bf d}$.
For aberrated or lossy imaging systems, information is lost going from ${\bf d}$ to ${\bf m}$ making ${\bf H}$ non-invertable, or leading to noise amplification. 

We therefore face an ill-posed inverse problem and instead seek the pseudo-inverse ${\bf H}^{(\rm PI)}$ that most faithfully recovers ${\bf d}\approx {\bf H}^{(\rm PI)} {\bf m}$.
For example, the expression
\begin{align}
{\bf H}^{(\rm PI)}(\alpha) &= \frac{1}{{\bf H}^\dagger {\bf H} + |\alpha|^2} {\bf H}^\dagger\label{eq:Tikhonov}
\end{align}
defines a Tikhonov pseudo-inverse~\cite{Tikhonov77}.
The eigenvalue expansion for ${\bf H}^\dagger {\bf H}$ indicates that $|\alpha|^2$ introduces a minimum eigenvalue into the denominator of Eq.~\eqref{eq:Tikhonov}, thereby avoiding singular behavior when any eigenvalue of ${\bf H}^\dagger {\bf H}$ becomes zero.
Wiener deconvolution in signal processing~\cite{Wiener1949, Orieux2010} is an example of a Tikhonov pseudo-inverse, where different values of $|\alpha|^2$ are associated with each eigenvalue of ${\bf H}^\dagger {\bf H}$.
Here, we derive a pseudo-inverse of the Tikhonov form by combining a noise model of the measurement process with a Bayesian prior for the density distribution, asserting that the distribution is confined in a compact region of space.

This paper is organized as follows: in Sec.~\ref{sec:TheoryBeerLambert} we present a basic description of light propagating through a dilute atomic cloud.
Next, in Sec.~\ref{sec:ImagingTechniques} we provide a unified description of imaging cold-atom clouds, and identify absorption and phase contrast imaging methods in suitable limits.
In Sec.~\ref{sec:Regularization}, we discuss our aberration correction algorithm and test it on simulated data.
In Sec.~\ref{sec:Microscope}, we describe our microscope for imaging $^{87}{\rm Rb}$ Bose-Einstein condensates (BECs) and detail our implementation of phase contrast imaging. 
Lastly, in Sec.~\ref{sec:ExpResults}, we apply our regularization method to experimental data and compare with existing techniques, and demonstrate the utility of our method by non-destructively imaging the thermal to BEC phase transition {\it in-situ}.

\section{Fundamentals of Light Wave-Matter Interactions}
\label{sec:TheoryBeerLambert}

The majority of ultracold atom measurements rely on images of light that has interacted with an atomic ensemble.
As such, in this section we summarize the theoretical description of laser light propagating along $\ez$ through a dilute atomic cloud: a nonpermeable dielectric medium.
We relate the absorption and phase shift of the incident laser to a fundamental quantity in ultracold atom experiments: the 2D column density $\rho_{\rm 2D}({\bf r}_\perp) = \int \rho(\mathbf{r}) dz$, where $\rho(\mathbf{r})$ is the 3D atomic density with spatial coordinates ${\bf r} = x\ex + y\ey + z\ez$ and transverse coordinates ${\bf r}_\perp = x\ex + y\ey$. 

\subsection{The Paraxial Helmholtz Equation} 
\label{Section:Paraxial}
We consider a monocromatic laser with wavelength $\lambda$, wavenumber $k_{0}$\,=\,$2\pi / \lambda$, and angular frequency $\omega_0 = c k_0$ propagating in a medium with 
complex relative permittivity  $\varepsilon(\mathbf{r})=\epsilon/\epsilon_0$. Here $c$ is the free-space speed of light; $\epsilon$ is the permittivity; and $\epsilon_0$ is the electric constant.
The optical electric field $\bm{\mathcal{E}}(\bf r)$ is described by the vectorial wave equation
\begin{equation}
	\nabla^{2} \bm{\mathcal{E}}({\bf r}) + k_{0}^{2} \varepsilon(\mathbf{r}) \bm{\mathcal{E}}(\bf r) = - \nabla [\bm{\mathcal{E}}({\bf r}) \cdot \nabla \ln\varepsilon(\mathbf{r})]. 
\label{Eqn:HelmholtzGen}
\end{equation} 
The right-hand side of Eq.~\eqref{Eqn:HelmholtzGen} can be omitted when $\varepsilon(\mathbf{r})$ is slowly varying. 
Since we consider an incident laser beam traveling along $\ez$, we isolate the $z$ derivative to obtain the scalar wave equation
\begin{eqnarray}
	-\frac{\partial^2 \mathcal{E}\mathbf{(r)}}{\partial z^2} = \left [\nabla^2_\perp + k_0^2 \right] \mathcal{E}\mathbf{(r)} +k_0^2 \chi({\bf r}) \mathcal{E}(\mathbf{r}),
\label{HelmholtzDirZ}
\end{eqnarray}
for each polarization. 
Here $\nabla_{\bot}^{2}$\,=\,$\partial^{2}/\partial x^{2} + \partial^{2}/\partial y^{2}$ is the transverse Laplacian and $\chi({\bf r}) = \varepsilon(\mathbf{r})-1$ is the relative susceptibility.
Next, we assume that the variations of the field along ${\bf r}_\perp$ are on a scale large compared to $\lambda$ and express the field as $\mathcal{E}({\bf r}_\perp,z) = { E}({\bf r}_\perp, z)e^{ik_0z}$ emphasizing the propagation axis $\mathbf{e}_z$.
Inserting this expression into Eq.~\eqref{HelmholtzDirZ} and making the paraxial approximation by dropping the $\partial^2 E ({\bf r}_\perp, z)/\partial z^2$ term, we obtain the paraxial Helmholtz equation
\begin{equation}
	-2 i k_0\frac{\partial {E}({\bf r}_\perp, z)}{\partial z} = [\nabla^2_\perp  +k_0^2 \chi({\bf r})] { E}({\bf r}_\perp, z),
\label{eqn:ParaxHelmholtz}
\end{equation}
describing the paraxial wave field ${ E}({\bf r}_\perp, z)$.
In free space, with $\chi=0$, Eq.~\eqref{eqn:ParaxHelmholtz} is exactly solved by the differential operator
\begin{align}
{\bf K}(\Delta z) &= \exp\left(i\frac{\nabla_\perp^2}{2 k_0} \Delta z \right),
\label{eqn:ParSpectral}
\end{align}
that transforms a field at position $z$ to position $z+\Delta z$ according to $E({\bf r}_\perp, z + \Delta z) = {\bf K}(\Delta z) E({\bf r}_\perp, z)$ for any $\Delta z$.
In the spectral domain ${\bf K}(\Delta z)$ is diagonal, allowing free-space propagation to be implemented by simple scalar multiplication.

By contrast, no general solution exists when $\chi({\bf r})\neq0$. However, for $\Delta z$ small compared to the depth of field (DoF) $d_{\rm dof} = 2 k_0/k_{\rm max}^2$, the operator
\begin{align}
{\bf R}(\Delta z) &= \exp \left[ i \frac{k_0}{2} \int_z^{z+\Delta z} \chi({\bf r})\mathrm{d}z\right],
\label{eqn:ParCoordinate}
\end{align}
approximately transforms the field a distance $\Delta z$ via $E({\bf r}_\perp, z +\Delta z) \approx {\bf R}(\Delta z) E({\bf r}_\perp, z)$.
Where $k_{\rm max}$ (bounded above by $k_0$) is the largest transverse wavevector in the detected optical field.
$k_{\rm max}$ is first set by the object plane field and then further limited by the NA of the imaging system (see Sec.~\ref{sec:Microscope}).
In the thin object limit $\delta z \ll d_{\rm dof}$, where $\delta z$ is the total thickness of the object, Eq.~\eqref{eqn:ParCoordinate} gives the field $E({\bf r}_\perp, z +\delta z)$ just following the object without further consideration.

To describe the propagation of $E({\bf r}_\perp, z)$ through extended objects, Eq.~\eqref{eqn:ParaxHelmholtz} can be evaluated numerically, for example with split-step Fourier techniques~\cite{Korpel86}.
For this purpose, we divide the evolution into spectral and coordinate steps~\cite{Putra2014}. 
The symmetrized expression
\begin{equation}
E\left({\bf r}_\perp, z +\Delta z\right)\approx\mathbf{K}\left(\Delta z/2\right)\mathbf{R}(\Delta z)\mathbf{K}\left(\Delta z/2\right)E\left({\bf r}_\perp, z\right)
\label{eqn:SSFM}
\end{equation}
is valid through second order in $\Delta z$, as can be readily derived from the Baker-Campbell-Hausdorff identity.

The optical field following the object $E_+$ travels through an imaging system to the image plane, where its time-averaged intensity $I_+({\bf r}) = c\epsilon_0\left|E_+({\bf r})\right|^2/2$ (not field) is detected by a charge-coupled device (CCD). The time-average results from the fact that a typical $\sim 10\ {\mu s}$ imaging time is vast compared to the $2\pi/\omega_0$ optical period.

\subsection{Depth of field effects} 
\label{Section:DOF_effects}

In this section, we consider extended objects for which the thin object limit is inapplicable.
A realistic object is present only in some compact domain from $z_-$ to $z_+$ along the axis of light propagation ${\bf e}_z$.
We divide the field into two components
\begin{equation}
{E_+({\bf r}_\perp, z)} \equiv E_0({\bf r}_\perp, z) + \delta E({\bf r}_\perp, z),
\label{eqn:FieldDivide}
\end{equation}
where $E_0({\bf r}_\perp, z)$ describes the field with no object [$I_0({\bf r}_\perp, z)$ is the associated intensity.] and therefore obeys the free space paraxial wave equation, and $\delta E({\bf r}_\perp, z)$ describes the light scattered by the object. 
We focus on the normalized scattered field 
\begin{equation}
	f({\bf r}_\perp, z) = \frac{\delta E({\bf r}_\perp, z)}{E_0({\bf r}_\perp, z)},
\label{eqn:Fscatt_Ratio}
\end{equation}
subject to the boundary condition $\delta f({\bf r}_\perp, z) =0$ for $z<z_-$.
When the DoF of $E_0({\bf r}_\perp, z)$ greatly exceeds the $\delta z= z_+-z_-$extent of the object~\footnote{Although an arbitrary field can have $k_{\rm max}$ up to $k_0$, a typical Gaussian probe beam has a large beam waist, giving small a $k_{\rm max}$ with a large DoF.}, $f({\bf r}_\perp, z)$ obeys
\begin{align*}
	i \frac{\partial f({\bf r}_\perp, z)}{\partial z} + \frac{1}{2 k_0}[\nabla^2_\perp  + k_0^2 \chi({\bf r}_\perp, z)] f({\bf r}_\perp, z) &= - \frac{k_0}{2} \chi({\bf r}_\perp, z),
\end{align*}
a  paraxial wave equation as in Eq.~\eqref{eqn:ParaxHelmholtz} with a source term.
In the limit of small $\chi({\bf r}_\perp, z)$ and $f({\bf r}_\perp, z)$, we obtain the first order approximate expression
\begin{align}
	i \frac{\partial f({\bf k}_\perp, z)}{\partial z} - \frac{1}{2 k_0} {k^2_\perp} f({\bf k}_\perp, z) &= - \frac{k_0}{2} \chi({\bf k}_\perp, z) 
\label{eqn:dEratio_Fourier}
\end{align} 
in the spectral domain~\footnote{We implicitly indicate Fourier transforms by a wavevector such as ${\bf k}_\perp$ as an argument.}.
This expression is exactly solved by
\begin{equation}
	f({\bf k}_\perp, z_+)  = \frac{i k_0}{2}  \int_{z_-}^{z_+} \chi({\bf k}_\perp, z) \exp\left[-i\frac{k_\perp^2}{2 k_0} (z_+ - z) \right] \mathrm{d}z.
\label{eqn:f_ExactSol}
\end{equation}

In the following, we consider an imaging system focused at $z=0$ and ask ``What infinitely thin object located at $z=0$ yields the same scattered field as an extended object does?"
This is answered by first finding $f({\bf k}_\perp, z_+)$  ($z>z_+$ it obeys the free space paraxial equation), then back-propagating $f({\bf k}_\perp, z_+)$ to $z=0$, finally giving
\begin{equation}
	f_{\rm{eff}}({\bf k}_\perp)  = \frac{i k_0}{2}  \int_{z_-}^{z_+} \chi({\bf k}_\perp, z) \exp\left(+i\frac{k_\perp^2}{2 k_0} z \right) \mathrm{d}z .
\label{eqn:f_eff_integral}
\end{equation}
Extending the bounds of integration to $\pm \infty$ converts the $z$ integral to a 1D Fourier transform with a wavevector $-{k}^2_\perp/2k_0$, leading to the final expression 
\begin{equation}
	f_{\rm{eff}}({\bf k}_\perp)  = \frac{i k_0}{2} \tilde \chi\left({\bf k}_\perp, \frac{k_\perp^2}{2 k_0}\right) \equiv \frac{i k_0}{2} \chi_{\rm{eff}}({\bf k}_\perp).
\label{eqn:f_eff}
\end{equation}
The tilde in $\tilde \chi({\bf k}_\perp, k_{\perp}^2/2 k_0)$ emphasizes that the $z$ index is Fourier transformed as well.
Here we interpret the field $f_{\rm{eff}}$ as resulting from an effective 2D susceptibility $\chi_{\rm{eff}}({\bf k}_\perp)$. 

In many cases of physical interest the 3D susceptibility can be expressed in the separable form $\chi({\bf r}) = Z(z)\times\chi_{\rm 2D}({\bf r}_\perp)$, where $Z(z)$ is a  normalized real valued transverse mode function.
In this case $\chi_{\rm eff}({\bf k}_\perp) = h_{\rm dof}({\bf k}_\perp)\chi_{\rm 2D}({\bf k}_\perp)$ where, anticipating the notation that will be used in Sect.~\ref{Section:Aberrations}, we define the DoF contrast transfer function $h_{\rm dof}({\bf k}_\perp) \equiv \tilde Z(k_z)$ in terms of the Fourier transformed mode function, with $\tilde Z(0) = 1$ implied by $Z$'s normalization.
Throughout this paper we will take $Z(z)$ to be symmetric, implying $h_{\rm dof}({\bf k}_\perp) = h_{\rm dof}(-{\bf k}_\perp)$ is real valued.

For the special case of a Gaussian mode function with $1/e$ width $w_z$, the DoF transfer function is
\begin{align}
	h_{\rm dof}({\bf k}_\perp) &= 
	\exp \left[ - \frac{1}{4} \left(\frac{w_z}{d_{\rm dof}}\right)^2
	\left(\frac{k_{\perp}}{k_{\rm{max}}}\right)^4\right].
\label{eqn:f_eff_2}
\end{align}
As a consequence the amplitude is suppressed for increasing $k_\perp$, but the phase is unaltered.
At $k_{\rm{max}}$ the suppression is $\exp \left[\left(-{w_z}/{2d_{\rm dof}}\right)^2\right]$, implying that there is negligible loss of information for objects appreciably thinner than the DoF, i.e.,  $w_z\ll d_{\rm dof}$.

\subsection{Atomic Susceptibility} 
\label{Section:Susceptibility}

For an ensemble of two-level atomic systems, the atom-light interaction is captured by the electric susceptibility
\begin{equation}
	\chi({\bf r})  =  \frac{\sigma_{0}}{k_0}\left[\frac{i-2 \bar\delta}{1+ \bar I({\bf r}) + 4 {\bar\delta}^2}\right] \rho(\mathbf{r}),
\label{eqn:ASuscept}
\end{equation}
where $\bar\delta = \delta / \Gamma$ 
is the normalized detuning from atomic resonance in terms of the detuning $\delta = \omega_0 - \omega_{\rm ge}$ and the natural atomic linewidth $\Gamma$; $\hbar\omega_{\rm ge}$ is the atomic transition energy; $\bar I({\bf r}) = I({\bf r})/I_{{\rm sat}}$ is the optical intensity in units of the saturation intensity $I_{{\rm sat}}$; and $\sigma_{0} = 6\pi/k_0^2$ is the resonant scattering cross-section. 

The atomic susceptibility $\chi({\bf r})$ is a complex quantity in which the real and imaginary parts result from distinct physical processes.
The real part derives from stimulated emission (i.e., forward scattering) resulting in a dispersive atomic medium with a density dependent index of refraction.
The imaginary part derives from spontaneous emission (i.e., nominally isotropic scattering) resulting in a density dependent absorption coefficient.
As a result, the optical field will be phase shifted and attenuated as it travels through the atomic cloud. 
We correspondingly express the field just after interacting with the atomic medium 
\begin{equation}
	E_+({\bf r}_\perp, z +\delta z) = e^{-\alpha({\bf r}_\perp) + i\phi({\bf r}_\perp)} E_0({\bf r}_\perp, z)
\label{eqn:Field_RealImgnry}
\end{equation}  
in terms of an absorption coefficient
\begin{align}
	\alpha({\bf r}_\perp) &= \frac{\sigma_0 \rho_{2\rm D}({\bf r}_\perp)}{2} \frac{1}{1+\bar I({\bf r}_\perp) + 4 \bar \delta^2} 
\label{eqn:Abs}
\end{align}
and a phase shift
\begin{align}
	\phi({\bf r}_\perp) &= -2 \bar\delta \alpha({\bf r}_\perp). 
\label{eqn:Phase}
\end{align}
These are both proportional to the optical depth 
\begin{align}
{\rm OD}({\bf r}_\perp) &\equiv -\ln\left[
\frac{I_+({\bf r}_\perp)}{I_0({\bf r}_\perp)}\right] 
\label{eqn:OD_Intensity}
\end{align}
via the relations
\begin{align}
\alpha({\bf r}_\perp) &= \frac{{\rm OD}({\bf r}_\perp)}{2} & {\rm and} && 	\phi({\bf r}_\perp) &=-\bar \delta {\rm OD}({\bf r}_\perp).
\label{eq:ABS_Phase_OD}
\end{align}
The 2D column density is related to the optical depth in terms of both the detuning and intensity
\begin{equation}
\sigma_0 \rho_{\rm 2D}({\bf r}_\perp) = \left[1 + 4\bar\delta^2 \right]{\rm OD}({\bf r}_\perp) + \bar I_0({\bf r}_\perp)\left[1 - e^{-{\rm OD}({\bf r}_\perp)}\right].
\label{eqn:OD_2Dcolumn}
\end{equation}
This expression shows that irrespective of how it was obtained, the optical depth serves to define the column density.
In the limit of small optical depth,  Eq.~\eqref{eqn:OD_2Dcolumn} reduces to
\begin{equation}
\sigma_0 \rho_{\rm 2D}({\bf r}_\perp) \approx \left[1 + \bar I_0({\bf r}_\perp) + 4\bar\delta^2 \right] {\rm OD}({\bf r}_\perp);
\label{eqn:Small_OD}
\end{equation}
this could result from any combination of low density, large detuning or high intensity.
For a spatially thin medium ($\delta z\ll d_{\rm dof}$) and imaging with low intensity laser light ($I_0 \ll I_{\rm sat}$) on resonance ($\bar\delta=0$), the optical depth following Eq.~\eqref{eqn:OD_2Dcolumn} is ${\rm OD}({\bf r}_\perp) = \sigma_0 \rho_{\rm 2D}({\bf r}_\perp)$.

\section{Imaging Techniques with Cold Atoms}
\label{sec:ImagingTechniques}
In this section, we describe two well-established imaging methods that are frequently employed in cold-atom experiments: phase contrast imaging (PCI) and absorption imaging (AI). 
We begin with the analysis of the general imaging scheme illustrated in Fig.~\ref{Schematic_PCI}, which includes a small phase shifter (phase dot) that is absent (i.e. gives $0$ phase shift) for AI. 

The object attenuates and diffracts the incident light, as described by Eq.~\eqref{eqn:Field_RealImgnry}, which can be re-expressed in terms of unscattered and scattered components. 
Using Eq.~\eqref{eqn:FieldDivide} this gives the object plane field
\begin{align}
	{E_+({\bf r}_\perp, z)} & \equiv E_0({\bf r}_\perp, z) + E_0({\bf r}_\perp, z)\left[e^{i \phi({\bf r}_\perp) - \alpha({\bf r}_\perp)}-1 \right].
	\label{eqn:Field_Sc_USc}
\end{align}
A phase dot shifts the optical phase of the unscattered light by $\theta$, giving
the image plane field $E^\prime_0({\bf r}_\perp, z) = E_0({\bf r}_\perp, z) \exp(i\theta)$, while leaving the scattered component unchanged. 
The resulting expression for the normalized image plane field after interacting with the atoms and the phase dot is
\begin{equation}
	\frac{E^\prime_+({\bf r}_\perp, z)}{E^\prime_0({\bf r}_\perp, z)} = 1 + e^{-i\theta}\left[e^{i \phi({\bf r}_\perp) - \alpha({\bf r}_\perp)}-1 \right].
\label{eqn:Field_PhDot_OD}
\end{equation}
Equation~\eqref{eq:ABS_Phase_OD} leads to the relation
\begin{equation}
	\frac{E^\prime_+({\bf r}_\perp, z)}{E^\prime_0({\bf r}_\perp, z)} = 1 + e^{-i\theta} \left\{ \exp\left[-\left(\frac{1}{2}+ i\bar\delta\right){\rm OD}({\bf r}_\perp)\right]-1\right\}
\label{eqn:NormField_PhDot_OD}
\end{equation}
between the normalized field and the optical depth.
Experimentally we detect the intensities $I^\prime_0({\bf r}_\perp)$ and $I^\prime_+({\bf r}_\perp)$, the image plane intensities associated with the object plane intensities $I_0({\bf r}_\perp)$ and $I_+({\bf r}_\perp)$.
Equation~\eqref{eqn:NormField_PhDot_OD} leads to the normalized signal
\begin{equation}
\begin{split}
	g^{\prime}_{\theta}({\bf r}_\perp) &= 2 \cos(\theta) - e^{-{\rm OD({\bf r}_\perp)}} - 1 + 2e^{-{\rm OD({\bf r}_\perp)}/2} \\
	& \times \left[\cos(\bar\delta {\rm OD({\bf r}_\perp)}) -\cos(\theta + \bar\delta {\rm OD({\bf r}_\perp)}) \right],
\end{split}
\label{eqn:g_General}
\end{equation}
where $g^{\prime}_{\theta}({\bf r}_\perp) \equiv 1 - {I_+^{\prime}({\bf r}_\perp)} / {I^\prime_0({\bf r}_\perp)}$. 
This noninvertible expression is applicable to both AI and PCI. 

In the following sections we derive the optical depth from this transcendental equation in limits appropriate for AI and PCI, and thereby leading to the column density through Eq.~\eqref{eqn:OD_2Dcolumn}.

\subsection{Phase Contrast Imaging} 
\label{Section:PCI}
In 1932, Frits Zernike invented PCI as a phase sensitive imaging method utilizing the nonuniform refractive index of an object to reveal features that are invisible in other imaging techniques~\cite{ZERNIKE1942686, ZERNIKE1942974}. 
Today, PCI has found application in various fields as a noninvasive {\it in-situ} imaging method~\cite{Oettle48, Fassett_PCI}.
In this section, we first introduce the basic principle of PCI and then derive the theoretical toolbox enabling a quantitative treatment of PCI in ultracold atom systems. 

\subsubsection{Principle of phase contrast imaging}
\label{SubSec:PCIgen}

\begin{figure}[t!]
	\begin{center}
	\includegraphics[width=3.3in]{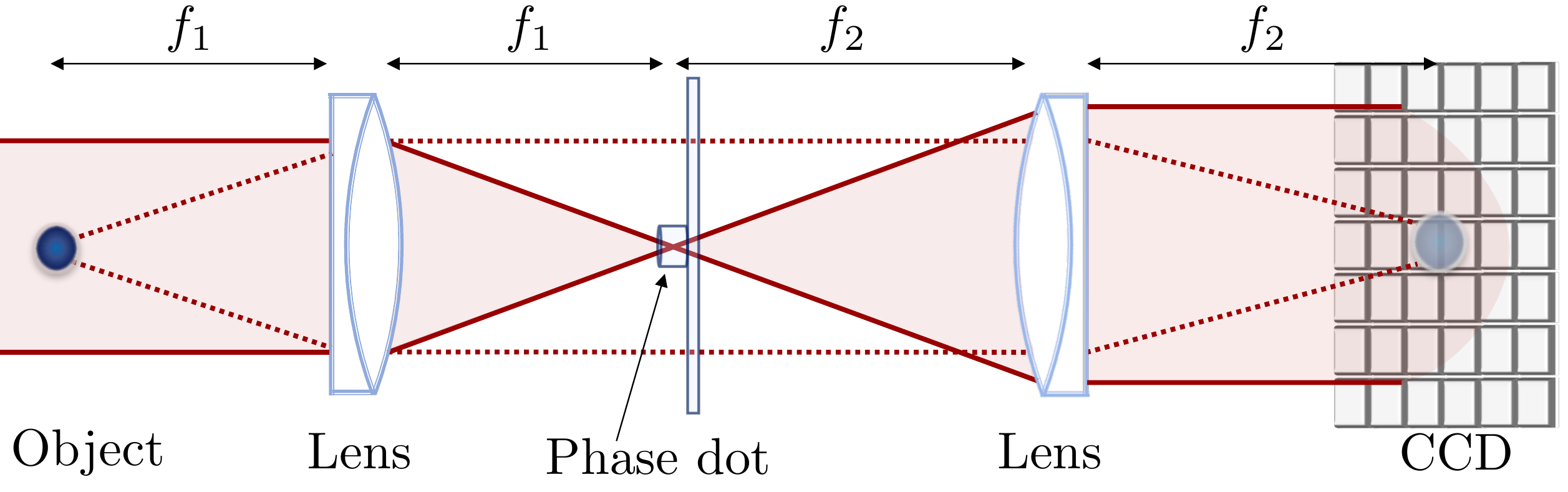}
	\end{center}
	\caption[Illustration of phase contrast imaging setup]{
	Schematic illustrating the principle of phase contrast imaging. 
	A refractive object scatters light from an incident probe laser into
	 two components: unscattered (solid) and scattered (dashed).
	 An objective lens placed a focal distance $f_1$ from the object spatially separates the two components at the back Fourier plane a distance $f_1$ from the lens.
	 The phase dot is positioned at the Fourier plane and predominately phase shifts the unscattered light passing through it. 
	 After the second lens with focal length $f_2$ the two components interfere in the image plane and a CCD records the resulting intensity.}
	\label{Schematic_PCI}
\end{figure}

PCI is an interferometric technique sensitive to the phase shift of light having propagated through an object. 
The extensive application of the technique stems from the elegant simplicity of the required instrumentation. 
By imprinting a position dependent phase shift $\phi({\bf r})$ on to the incident field, the object diffracts part of that light (see Fig.~\ref{Schematic_PCI}). 
PCI can be understood as an interferometer in which the unscattered component is the reference beam (the local oscillator) and the scattered component carries information about the object. 
These two components share the same optical path making PCI  robust against vibrations in the imaging system.

Both components are collected by an imaging lens that is positioned at its focal distance $f_1$ from the object. 
The unscattered light comes to an intermediate focus at the back Fourier plane of the lens, spatially separating the scattered and the unscattered components.
A small dielectric dot (phase dot) just larger than the focused unscattered beam is positioned at the Fourier plane as shown in Fig.~\ref{Schematic_PCI}. 
The phase dot shifts the phase of the unscattered light by $\theta$ but leaves the scattered component unchanged~\cite{Ketterle99}.
A second imaging lens with focal length $f_2$ forms an image plane where scattered and the unscattered components of the wave field interfere. 
At the image plane intensity is detected, with an overall magnification $M = f_2/f_1$.

\subsubsection{Phase Contrast Imaging with Ultracold Atoms} \label{SubSec:PCI_Atoms}
The PCI intensity encodes information about the object-plane phase from which we extract the optical depth of the atomic cloud.
In the limit of large laser detuning, where PCI is typically applied, we neglect absorption because $\alpha \ll \phi$.
In this limit, Eq.~\eqref{eqn:g_General} for the normalized intensity reduces to
\begin{equation}
g^{\prime}_{\theta}({\bf r}_\perp) = 2\left\{ \cos\theta + \cos\phi({\bf r}_\perp) - \cos \left[\theta + \phi({\bf r}_\perp)\right]- 1 \right\}.
\label{eqn:PCI_Int}
\end{equation}
In the limit of small phase shift (i.e.,  $\phi\ll 1$) the normalized PCI intensity
\begin{equation}
	g^{\prime}_{\theta}({\bf r}_\perp) \approx 2 \phi({\bf r}_\perp) \sin\theta \xrightarrow{\theta=\pi/2} 2\phi({\bf r}_\perp)  
	\label{eqn:PCI_Piover2}
\end{equation}
is linearly proportional to the phase shift imparted by the object and maximized for $\theta=\pi/2$.
PCI yields an increased sensitivity for weak phase objects compared to other dispersive imaging methods~\cite{Ketterle99}. Lastly, we obtain the optical depth
\begin{equation}
{\rm OD}_{\rm PCI}({\bf r}_\perp,\bar \delta \gg 1) = \frac{1}{2\bar \delta}g^{\prime}_{\theta=\pi/2}({\bf r}_\perp)
\label{eqn:OD_PCI_piOver2}
\end{equation}
using Eq.~\eqref{eq:ABS_Phase_OD}.

The minimally destructive nature of PCI measurement becomes evident for large detunings.
In the limit $\delta \gg \Gamma$ while the phase shift imparted by the atomic system is $\propto 1/\delta$, the spontaneous emission rate is $\propto 1/\delta^2$. 
As a result, atom loss due to radiation pressure becomes negligible.
Hence, in ultracold atom experiments PCI is typically employed to non-destructively image high column density atomic clouds {\it in-situ} and at large detuning~\cite{Ketterle96, Ketterle97, Anderson2001}. 

\subsection{Absorption Imaging} 
\label{Section:Absorption Imaging}
AI of ultracold atoms usually employs resonant or near-resonant laser light, i.e., $|\delta|\lesssim\Gamma$, where the spontaneous scattering of photons creates a shadow in the outcoming light wave. 
We measure this shadow and infer the column density of the object from the resultant images.

The imaging system in Fig.~\ref{Schematic_PCI}, introduced in the context of PCI, is applicable to AI provided the phase dot is removed.
Accordingly we apply the formalism in Equ.~\eqref{eqn:g_General}, with $\theta =0$. 
The on resonance ($\bar \delta = 0$) optical depth in terms of the detected normalized intensity in the image plane is
\begin{equation}
{\rm OD}_{\rm AI}({\bf r}_\perp) = -\ln\left[1-g^{\prime}_{\theta=0}({\bf r}_\perp)\right].	
\label{eqn:OD_AI}
\end{equation}

\subsubsection{Partial transfer absorption imaging} 
\label{Section:PTAI}

The high-optical density of most BECs prevents the direct observation of their density {\it in-situ} using standard AI.
Dense clouds absorb the vast majority of the incident probe laser, leading to ODs greatly in excess of 1.
This compromises the SNR, and in practice the OD saturates around 4.
Although detuning the probe beam reduces the atomic cross-section, the cloud behaves like a gradient index lens leading to imaging distortions~\cite{Ketterle96, Reinaudi2007}. 
Dispersive imaging techniques such as PCI~\cite{Ketterle99}, dark-field imaging~\cite{Ketterle96} and Faraday imaging~\cite{Gajdacz2013} can operate at large detuning $\delta \gg \Gamma$, where phase shifts are small and lensing effects are thereby reduced.
Partial transfer absorption imaging (PTAI) is an alternate approach for imaging high density atomic ensembles.
In PTAI, an RF or microwave pulse transfers a fraction of the atoms from a dark state to a bright detection state where they are absorption imaged~\cite{Freilich1182, Campbell2012}.
In this way, PTAI mitigates large OD effects and can yield minimally destructive repeated images of the same atomic system~\cite{Seroka2019}.

In our specific experiment PTAI has additional quantum projection noise effects. 
For deeply degenerate interacting BECs, number fluctuations are greatly suppressed~\cite{Schley2013}; the RF/microwave transfer process in PTAI then leads to enhanced atom shot noise similar to how a beam splitter introduces vacuum port noise in quantum optics. 
In Sec.~\ref{sec:ExpPupilFunc} we utilize this fact when measuring the pupil function of our microscope. 

\subsection{Aberrations} 
\label{Section:Aberrations}

Here we model aberrations as a Fourier pupil function that both phase-shifts and attenuates the optical field as a function of wavevector.
Importantly, this model can only treat aberrations where the PSF---the magnitude squared of the impulse response function---is the same everywhere in the observed field of view.

Motivated by our introduction of regularization, we introduce the forward transfer function
\begin{align}
h_E({\bf k}_\perp) &= e^{-\gamma(\mathbf{k}_\perp) + i\beta(\mathbf{k}_\perp)} \label{eq:transfer},
\end{align}
describing the navigation of fields through our imaging system (neglecting the PCI phase dot) via $E^\prime_{+/0}({\bf k}_\perp) = h_E({\bf k}_\perp) E_{+/0}({\bf k}_\perp)$. 
Here $\gamma({\bf k}_\perp)$ describes attenuation and $\beta({\bf k}_\perp)$ describes phase shifts.
Even ideal imaging systems will have contributions from these terms.
For example, defocus will contribute a quadratic $\beta\propto k_\perp^2$ term, and the NA limits the maximum accepted wavevector to $k_{\rm NA} = {\rm NA} \times k_0 $, implying $\gamma(\mathbf{k}_\perp)\rightarrow\infty$ for $|\mathbf{k}_\perp| > k_{\rm NA}$.

In our discussion of PCI, we assumed that the field $E_0({\bf r}_\perp)$ with the atomic ensemble absent is slowly varying and therefore contains Fourier components only near $\mathbf{k}_\perp=0$. Thus following the imaging system it is transformed to $E^\prime_0({\bf r}_\perp) = h_E(0) E_0({\bf r}_\perp)$.
Including the impact of the phase dot as well as DoF effects introduced in Sect.~\ref{Section:DOF_effects}, we arrive at the image-plane field ratio
\begin{align}
f^\prime({\bf k}_\perp) &= h_{\rm dof}({\bf k}_\perp)\frac{h_E({\bf k}_\perp)}{h_E(0)} e^{-i\theta} \delta E({\bf k}_\perp).\label{eq:PCItransfer}
\end{align}
Linearizing Eq.~\eqref{eqn:Field_Sc_USc} connects the image-plane field ratio to the optical depth via
\begin{align}
f^\prime({\bf k}_\perp) &= h_{\rm tot}({\bf k}_\perp) {\rm OD}({\bf k}_\perp),\label{eq:field_transfer}
\end{align} 
in terms of the total transfer function
\begin{align}
h_{\rm tot}({\bf k}_\perp) &= \sqrt{\frac{1}{4}+\bar\delta^2} h_{\rm dof}({\bf k}_\perp) \frac{h_E({\bf k}_\perp)}{h_E(0)} e^{i(\varphi-\theta)}
\label{eq:augtransfer}.
\end{align}
Here $\varphi$, defined via $\tan\varphi = 2\bar\delta$, describes the complex angle associated with the atomic susceptibility.
We see that the $\mathbf{k}_\perp=0$ contributions to the pupil function have no impact, implying that any inferred dc component to the pupil function only results from detuning and the PCI phase shift as parameterized by $\varphi-\theta$.

Expressing this ratio as an intensity in coordinate space and converting back to the spectral domain gives
\begin{align}
 g^\prime({\bf k}_\perp) &=\overbrace{ \left[h_{\rm tot}({\bf k}_\perp) + h_{\rm tot}^*(-{\bf k}_\perp)\right]}^{h({\bf k}_\perp)}  {\rm OD}(\mathbf{k}_\perp) \label{Eq:GeneralContrastTransfer},
\end{align}
where the quantity in square brackets is the contrast transfer function that encodes the optical depth as a change in fractional intensity.

This expression takes on a more conventional form when $\gamma$ and $\beta$ are expressed in terms of their symmetric and anti-symmetric contributions, i.e., $\gamma_\pm({\bf k}_\perp) = [\gamma({\bf k}_\perp) \pm \gamma(-{\bf k}_\perp)]/2$, and making the reasonable assumption of symmetric attenuation ($\gamma_- = 0$).
Then we obtain
\begin{align}
  h({\bf k}_\perp) =& \sqrt{1+4\bar\delta^2} h_{\rm dof}({\bf k}_\perp) e^{-\gamma_+({\bf k}_\perp) + i \beta_-({\bf k}_\perp)} \nonumber\\
&\times \cos[\beta_+({\bf k}_\perp) + \varphi-\theta] \label{Eq:SimplifiedContrastTransfer}.
\end{align}
For absorption imaging ($\theta=0$) of thin objects [$h_{\rm dof}({\bf k}_\perp) = 1$] with no loses ($\gamma=0$) and a quadratic phase shift $\beta=z k_\perp^2 / 2 k_{0}$, we arrive at the well-known result $h({\bf k}_\perp) = \cos(z k_\perp^2 / 2 k_{0}) + 2\bar\delta \sin(z k_\perp^2 / 2 k_{0})$, which results from defocus by a distance $z$ [see Eq.~\eqref{eqn:ParSpectral}]~\cite{Turner2005,Perry2021}. Furthermore, our result shows that up to an overall sign far detuned PCI with $|\varphi|=|\theta|=\pi/2$ obeys the same CTF as resonant AI.

These pupil functions can be calibrated using the fluctuations $\delta {\rm OD}({\bf r}_\perp) \equiv {\rm OD}({\bf r}_\perp) - \langle {\rm OD}({\bf r}_\perp) \rangle$, where $\langle\cdots\rangle$ denotes the average over an ensemble of images of cold atoms~\cite{Hung_2011}.
Assuming spatially uncorrelated density correlations, i.e., $\langle \delta {\rm OD}({\bf r}) \delta {\rm OD}({\bf r}^\prime) \rangle \propto \delta^{(3)}({\bf r}-{\bf r}^\prime)$, where $\delta^{(3)}({\bf r}_\perp)$ denotes the 3D Dirac delta function, the power spectral density is
\begin{align}
\langle |\delta {\rm OD}({\bf k}_\perp) |^2 \rangle \propto &\  e^{-2\gamma_+({\bf k}_\perp)}\bigg\{\cosh[2\gamma_-({\bf k}_\perp)]\label{Eq:Pupil_PSD}\\
&+ h_{\rm dof}({\bf k}_\perp)\cos[2\beta_+({\bf k}_\perp) + 2(\varphi-\theta)]\bigg\}\nonumber.
\end{align}
This signal is sensitive to all components of the pupil function except $\beta_-$.
In Sec.~\ref{sec:ExpPupilFunc}, we use this signal obtained at a range of image planes to extract low noise maps of the pupil function.

\subsection{Signal-to-noise ratio} 
\label{sec:SNR}
In this section we compare the SNR of PCI and AI.
In our measurements, we detect probe pulses of duration $\Delta t$ on a CCD sensor of square pixel size $\Delta x$ and quantum efficiency $\eta$. 
The intensity at pixel coordinates $\bf i$ is $I_{\bf i} = N_{\bf i} I_{\rm pe}$, where $N_{\bf i}$ is the number of photo-electrons and $I_{\rm pe} = {\hbar\omega_0}/{\eta A \Delta t}$ is the intensity required to generate a single photo-electron given the single-photon energy $\hbar\omega_0 = c \hbar k_0$. 
In a single experimental shot, our measurement techniques employ three images that yield (1) $I_{+,{\bf i}}$ of the probe in the presence of atoms, (2) $I_{0,\bf i}$ of the probe field without the atoms, (3) $I_{D, {\bf i}}$ with no probe light. 
For the remainder of the manuscript, we will omit the prime notation that distinguishes the image plane from the object plane.
We subtract $I_{D,{\bf i}}$ from $I_{+,{\bf i}}$ and $I_{0,{\bf i}}$ to eliminate any baseline from background illumination.

In bright field detection techniques, photon shot noise is the dominant source of noise, thereby we neglect other sources of technical noise such as dark current and read noise.
Photon counting can be modeled as a classical Poisson process where individual photon detections are treated as independent events with an uncorrelated temporal distribution. 
Photon shot noise (more specifically the shot noise of the detected photo-electrons) explains the width of this distribution, which has its variance equal to its mean. 
We model each detected image $I_{\bf i} = \langle I_{\bf i}\rangle+\delta I_{\bf i}$ as the sum of its mean $\langle I_{\bf i}\rangle$ and measurement noise $\delta I_{\bf i}$ (we will only consider zero mean random variables, i.e., $\langle\delta I_{\bf i}\rangle = 0$). 
Then the spatially uncorrelated photon shot noise is described by
\begin{equation}
\langle\delta I_{\bf i}\delta I_{\bf i^\prime} \rangle = \delta_{{\bf i},{{\bf i}^\prime}} I_{\rm pe} \langle  I_{\bf i} \rangle,
\label{Eq:PSNoise}
\end{equation}
where $\delta_{{\bf i},{\bf i}^\prime}$ is the Kronecker $\delta$ function.
Next we consider the noise in the fractional intensity $g_\theta$. 
In practice, we construct the background image $I_{0,{\bf i}}$ by averaging many images of the probe beam with no atoms present, and as a result it contributes negligible photon shot noise.
With this assumption and following Eq.~\eqref{Eq:PSNoise}, the noise in the fractional intensity is 
\begin{equation}
\langle\delta g_{\bf i} \delta g_{\bf i^\prime}\rangle = \delta_{{\bf i},{\bf i^\prime}} \frac{I_{\rm pe}}{I_{0,\bf i}}\left[1-\langle g_{\bf i}\rangle\right].
\label{eq:gNoise}
\end{equation}

Assuming that both the phase shift and OD are small, the noise variance of the OD deduced from PCI using Eq.~\eqref{eqn:OD_PCI_piOver2} is
\begin{equation}
\langle\delta {\rm OD}_{\bf i} \delta {\rm OD}_{\bf i^\prime} \rangle_{\rm PCI} = \delta_{{\bf i},{\bf i^\prime}} \frac{1}{4\bar\delta^2}\langle\delta g_{\bf i}^2\rangle.
\label{PCI_Noise}
\end{equation} 
For AI using Eq.~\eqref{eqn:OD_AI} noise variance is 
\begin{equation}
\langle\delta {\rm OD}_{\bf i} \delta {\rm OD}_{\bf i^\prime} \rangle_{\rm AI} = \delta_{{\bf i},{\bf i^\prime}} \frac{\langle\delta g_{\bf i}^2\rangle}{[1-\langle g_{\bf i}\rangle]^2}.
\label{AI_Noise}
\end{equation}
Together these expressions show that near resonance the SNR of AI  exceeds that of PCI, while far from resonance PCI has the larger SNR~\cite{Perry2021}.
In addition, the noise variance for AI diverges at large optical depth (where $\langle g_{\bf i}\rangle\rightarrow 1$) because the fractional photon shot noise increases with increasing absorption; this emphasizes the importance of PCI or PTAI for large OD systems.

Comparing the expressions for PCI and AI, we see that for fixed $I_{0,\bf i}$ (fixed back-action on atoms) the noise variance in PCI is lower by a factor of $\bar\delta$ compared to that of AI for large detuning and small absorption, i.e., low optical depth. 
This implies that AI cannot be a back-action limited measurement in this limit. 

\section{Regularization}
\label{sec:Regularization}

We consider the general inversion problem where the linear operator ${\bf H}$ describes a forward transformation to the measurement basis described by vectors ${\bf m}$, according to ${\bf m} = {\bf H}\ {\bf d}$, where we read ${\bf m}$ as the measurement outcome and ${\bf d}$ as the desired data.
Our approach follows a Bayesian line of reasoning, where we include a pair of priors and seek the most likely vector ${\bf d}$ given these priors.

\subsection{Bayesian framework}
\label{sec:BayesFramework}

Before moving forward, we introduce a Gaussian prior distribution function
\begin{align}
P_d({\bf d} ; {\bf p}, {\boldsymbol \Delta}) &\propto \exp\left[- \frac{({\bf d} - {\bf p})^\dagger {\boldsymbol \Xi}^{-1}   ({\bf d} - {\bf p})}{2}  \right],
\label{eq:prior}
\end{align}
giving the probability of finding the data vector ${\bf d}$ conditioned on knowing a prior ${\bf p}$ with confidence expressed by the covariance matrix ${\boldsymbol \Xi}$.
The diagonal entries of the covariance matrix $\Xi_{jj} = \xi_{j}^2$ derive from the conventional single-sigma uncertainties $\xi_j$.
An analogous distribution $P_m({\bf m}_0 ; {\bf m}, {\boldsymbol \Sigma})$ applies for measurements, giving the probability that the ``true'' measurement outcome was ${\bf m}_0$ conditioned on having observed ${\bf m}$ and knowing the covariance matrix ${\boldsymbol \Sigma}$, with diagonal entries $\sigma^2_i$.

By combining these expressions we obtain 
\begin{align}
P({\bf d}) &\propto P_d({\bf d} ; {\bf p}, {\boldsymbol \Xi}) \times P_m({\bf H}\ {\bf d} ; {\bf m}, {\boldsymbol \Sigma}),
\end{align}
the probability of finding the data vector ${\bf d}$, with forward transform ${\bf H}\ {\bf d}$, conditioned on both ${\bf p}$ and ${\bf m}$.
Here we select the most likely ${\bf d}$ as our pseudo-inverse, i.e., we employ maximum likelihood estimation.

By taking $-2\ln P({\bf d})$ we recast the inversion problem as a minimization problem with the quadratic objective function
\begin{align}
E =& \alpha^2 ({\bf d} - {\bf p})^\dagger \bar{\boldsymbol \Xi}^{-1}   ({\bf d} - {\bf p})  \label{eq:objective}\\
&+ ({\bf H}\ {\bf d} - {\bf m})^\dagger \bar{\boldsymbol \Sigma}^{-1}   ({\bf H}\ {\bf d} - {\bf m}).\nonumber
\end{align}
Here we introduced normalized covariance matrices $\bar{\boldsymbol \Xi} = {\boldsymbol \Xi} / \xi^2_{\rm max}$ and $\bar{\boldsymbol \Sigma} = {\boldsymbol \Sigma}/\sigma^2_{\rm min}$, where $\xi^2_{\rm max}$ is the largest eigenvalue of ${\boldsymbol \Xi}$; $\sigma^2_{\rm min}$ is smallest eigenvalue of ${\boldsymbol \Sigma}$; and $\alpha^2 = \sigma^2_{\rm min} / \xi^2_{\rm max}$ will function as a regularization parameter. 
The first term in Eq.~\eqref{eq:objective} describes the uncertainty-weighted difference between the prior ${\bf p}$ and the reconstruction ${\bf d}$, and the second term measures the uncertainty-weighted difference between the measurements ${\bf m}$ and the prediction of the reconstruction ${\bf H}\ {\bf d}$.

The objective function can be simplified by making use of the Cholesky decomposition, where the symmetric covariance matrices are expressed as $\bar {\boldsymbol\Sigma} = {\bf C}_\Sigma {\bf C}_\Sigma^\dagger$ and $\bar {\boldsymbol\Xi} = {\bf C}_\Xi {\bf C}_\Xi^\dagger$.
This leads to the simplified objective function 
\begin{align}
E &= |{\bf J}{\bf d}^\prime  - {\bf m}^\prime|^2 + \alpha^2 |{\bf d}^\prime-{\bf p}^\prime|^2 \label{eq:objective_simple}
\end{align}
in terms of a new operator ${\bf J} = {\bf C}_\Sigma^{-1} {\bf H} {\bf C}_\Xi$, and new vectors ${\bf m}^\prime = {\bf C}_\Sigma^{-1} {\bf m}$, ${\bf d}^\prime={\bf C}_\Xi^{-1}{\bf d}$ and  ${\bf p}^\prime={\bf C}_\Xi^{-1}{\bf p}$.

Since Eq.~\eqref{eq:objective_simple} is a quadratic form it has a unique minimum, which we obtain by setting the gradient
\begin{align}
{\boldsymbol \nabla}_{{\bf d}^\prime} E &= 2\left[\left(\alpha^2+ {\bf J}^\dagger {\bf J}\right) {\bf d}^\prime - {\bf J}^\dagger {\bf m}^\prime - \alpha^2{\bf p}^\prime\right] \label{eq:gradient}
\end{align}
equal to zero, where ${\boldsymbol \nabla}_{{\bf d}^\prime}$ is the gradient with respect to the ${\bf d}^\prime$ vector.
This gives the root
\begin{align}
{\bf d}^\prime_0 &= \left(\alpha^2+ {\bf J}^\dagger {\bf J}\right)^{-1}\left({\bf J}^\dagger {\bf m}^\prime + \alpha^2{\bf p}^\prime\right) \\
&\rightarrow \left(\alpha^2+ {\bf J}^\dagger {\bf J}\right)^{-1} {\bf J}^\dagger {\bf m}^\prime\label{eq:MinErrorZero}
\end{align}
where in the second line we selected the ${\bf p}^\prime=0$ null prior, thereby replicating the generic Tikhonov form presented in Eq.~\eqref{eq:Tikhonov}.

\subsection{Specific implementation}

Having employed a standard Bayesian framework to obtain a maximum-likely reconstruction, we now specialize to our imaging application. 

Our method uses this framework by adding new information: {\it outside} some window no atoms exist, but the atomic distribution within that window is completely unknown.
We thereby accept the ${\bf p} = 0$ prior outside the window by setting $\xi_j\rightarrow0$ in that region, and reject the prior inside the window by setting $\xi_j=1$ with $\alpha\ll 1$, implying that ${\boldsymbol \Xi}$ is diagonal in the final spatial basis.
In principle ${\boldsymbol \Sigma}$ includes all known sources of uncertainty: in our case only photon-shot noise in the detection system is significant, making ${\boldsymbol \Sigma}$ diagonal in the initial detection basis.
Lastly, we constrain our implementation to imaging imperfections described by Eq.~\eqref{Eq:GeneralContrastTransfer}, giving a forward transfer function $h_{\bf k}$ that is diagonal in the spectral basis.

Typical images are on the scale of $\approx10^3\times10^3$ pixels and therefore reside in a $\approx10^6$ dimensional vector space.
Since the resulting $\approx10^6\times10^6$ matrices in Eq.~\eqref{eq:MinErrorZero} are too large to manipulate  directly with today's desktop computers, in the following we describe implementations that do not require their explicit construction.
In addition, Appendix~\ref{app:grid} discusses further considerations involved in selecting a real-space grid large enough for artifact free reconstruction.
In general, padding the measured image ${\bf m}$ may be required.

\subsubsection{Spectral Tikhonov from uniform uncertainties}
\label{subsec:Tikhonov}

In the special case of uniform uncertainties---with $\bar\Sigma_{{\bf i}_1,{\bf i}_2} = \bar\Xi_{{\bf i}_1,{\bf i}_2} = \delta_{{\bf i}_1,{\bf i}_2}$, and $\alpha = \sigma/\xi$---it is natural to work in the spectral basis where ${\bf H}$ is diagonal and Eq.~\eqref{eq:MinErrorZero} reduces to 
\begin{align}
d_{\bf k} &=  \frac{h^*_{\bf k}}{\alpha^2 + |h_{\bf k}|^2} m_{\bf k}. \label{eq:Tikhonov_2}
\end{align}
This special-case expression is again of the Tikhonov form, but by contrast to the general solution in Eq.~\eqref{eq:MinErrorZero} it is diagonal in the spectral basis, making its deployment straightforward.
In practice, the regularization parameter $\alpha$ is empirically chosen and this inversion approach has been previously used to correct for the quadratic order aberrations resulting from defocus in cold-atom systems~\cite{Turner2005,Wigley2016a,Perry2021} as well as electron microscopy of biological systems~\cite{Penczek1997}.

\subsubsection{Ad hoc convolution approximation}
\label{subsec:adhoc}

Motivated by the simplicity of Eq.~\eqref{eq:Tikhonov_2}, we now derive an approximation to Eq.~\eqref{eq:MinErrorZero} that can still be implemented by multiplication in the spectral basis.

We again assume uniform detection uncertainties, but now allow $\bar{\bf \Xi}$ to be a window function which is diagonal in real space (and therefore implemented by a convolution in the spectral basis via the Fourier convolution theorem).
In the following discussion we use explicit summations rather than linear-algebra notation for an unambiguous presentation.
These assumptions lead to the simplification $\sum_{{\bf k}_2,{\bf k}_3}\bar\Xi_{{\bf k}_1 {\bf k}_2} H_{{\bf k}_2 {\bf k}_3} m_{{\bf k}_3}= \sum_{{\bf k}_2} \bar\Xi_{{\bf k}_1 - {\bf k}_2} h_{{\bf k}_2} m_{{\bf k}_2}$, allowing zero-gradient condition to be written as
\begin{align*}
    \sum_{{\bf k}_2}\left(\alpha^2 \delta_{{\bf k}_1 {\bf k}_2} +  \bar \Xi_{{\bf k}_1-{\bf k}_2} |h_{{\bf k}_2}|^2\right)  d_{{\bf k}_2} = \sum_{{\bf k}_2} \bar \Xi_{{\bf k}_1-{\bf k}_2} h^*_{{\bf k}_2} m_{{\bf k}_2}.
\end{align*}
We then make the ad hoc approximation of pulling $d_{{\bf k}}$ outside of the convolution, giving the simplified result
\begin{align}
    d_{{\bf k}} &\approx \frac{\sum_{{\bf k}_1} \bar \Xi_{{\bf k}-{\bf k}_1} h^*_{{\bf k}_1} m_{{\bf k}_1}}{\alpha^2  +  \sum_{{\bf k}_1} \bar \Xi_{{\bf k}-{\bf k}_1} |h_{{\bf k}_1}|^2}.\label{eq:SelfBayesianFinal}
\end{align}
The intuition behind this expression is that any zeros in the denominator are lifted by convolving with the Fourier transform of the window function---a smoothing process---thereby providing a form of regularization even for $\alpha=0$.

\begin{algorithm}[tbhp]
\footnotesize
\DontPrintSemicolon
\SetAlgoCaptionLayout{leftalign} 
\label{alg:adhoc}
\caption{Ad hoc approximation.}
\KwData{\;
\qquad $m_x$: Measured vector\;
\qquad $H_k$: Forward transform\;
\qquad $\bar\xi_x$: Normalised prior uncertainties\;
\qquad $\alpha$: Regularization parameter}
\KwResult{\;
\qquad $d_x$: Data vector 
}
\tcp{Compute numerator}
$d_k =  {\rm FT}_k(\bar\xi^2_x \  {\rm IFT}_x( H^*_{k^\prime} \ {\rm FT}_{k^\prime}(m_{x^\prime})))$\;
\tcp{Divide by denominator}
$d_k\ /\!= \alpha^2 + {\rm FT}_k(\bar\xi^2_x \  {\rm IFT}_x( |H_{k^\prime}|^2 ))$\;
$d_x = {\rm IFT}_x( d_k)$
\end{algorithm}

Algorithm~\ref{alg:adhoc} outlines the computational steps to implement the ad hoc convolution approximation.
In this pseudo-code, the Fourier transform ${\rm FT}_k(m_x)$ indicates that the resulting vector will have the momentum index $k$.
Expressions such as $\bar\Xi_x \ {\rm IFT}_x( |H_k|^2 )$ describe element-by-element multiplication and do {\it not} follow the Einstein summation convention, which would contract this quantity to a scalar. 
We evaluate the required convolutions via the Fourier transform-convolutions theorem, and hence Algorithm~\ref{alg:adhoc} does not require the explicit construction of large matrices. 

\subsubsection{Full method}\label{subsec:full}

In the full evaluation of Eq.~\eqref{eq:MinErrorZero}, we employ a conjugate-gradient algorithm~\cite{Press2007}, an efficient method that can be implemented without explicit construction of large matrices. Appendix~\ref{app:series} details a convergent infinite series expansion of Eq.~\eqref{eq:MinErrorZero}.
However this approach yielded poor performance compared to conventional numerical methods and we did not use it.

\begin{algorithm}[tbhp]
\footnotesize
\DontPrintSemicolon
\SetAlgoCaptionLayout{leftalign} 
\label{alg:conjugate gradient}
\caption{Conjugate gradient implementation solving $0 = {\bf Q} {\bf d} - {\bf b}$.
Here ${\bf Q} = \alpha^2 + \bar{\boldsymbol\Xi} {\bf H}^\dagger\bar{\boldsymbol\Sigma}^{-1}{\bf H}$ and ${\bf b} = \bar{\boldsymbol\Xi} {\bf H}^\dagger\bar{\boldsymbol\Sigma}^{-1} {\bf m}$. 
This algorithm assumes that $\bar{\boldsymbol\Xi}$ and $\bar{\boldsymbol\Sigma}$ are diagonal matrices with entries given by the vectors $\bar\xi_j^2$ and $\bar\sigma_j^2$ respectively. }
\KwData{\;
\qquad $m_x$: Measured vector\;
\qquad $H_k$: Forward transform\;
\qquad $\bar\xi_x$: Normalised prior uncertainties\;
\qquad $\bar\sigma_x$: Normalised measurement uncertainties\;
\qquad $\alpha$: Regularization parameter\;
\qquad $L$: Number of iterations
}
\KwResult{\;
\qquad $d_x$: Data vector 
}
\tcp{Initialize algorithm}
$b_x = \bar\xi^2_x\ {\rm IFT}_x(H^*_k\ {\rm FT}_k(\bar\sigma_{x^\prime}^{-2}\  m_{x^\prime}))$\;
$r_x = p_x = b_x$\;
$\epsilon = {\bf r}^\dagger {\bf r}$\;
\tcp{Implement algorithm}
\While{$L > 0$}
{
\qquad \tcp{Precompute ${\bf Q} {\bf p}$}
\qquad $Q^{p}_x = \bar \xi_x^2\ {\rm IFT}_x(H^*_k\ {\rm FT}_k(\bar\sigma_{x^\prime}^{-2}\ {\rm IFT}_{x^\prime} (H_{k^\prime}\ {\rm FT}_{k^\prime}(p_{x^{\prime\prime}}))))$\;
\qquad $Q^{p}_x\ +\!= \alpha^2 p_x$\;
\qquad $\gamma = \epsilon / ({\bf p}^\dagger {\bf Q}^{p})$\;
\qquad $d_x\ +\!=\gamma p_x$\;
\qquad $r_x\ -\!=\gamma Q^p_x$\;
\qquad $\epsilon^\prime = {\bf r}^\dagger {\bf r}$\;
\qquad $\beta = \epsilon^\prime / \epsilon$\;
\qquad $p_x = r_x + \beta p_x$\;
\qquad $L\ -\!=1$\;
}
\end{algorithm}

Algorithm~\ref{alg:conjugate gradient} charts our conjugate gradient approach implementation. Using this method the objective function in Eq.~\eqref{eq:objective} converges to within $\approx0.1\%$ of its asymptomatic value within 50 iterations.
We also implemented an adaptive step size gradient descent method with similar performance, but added complexity.
Therefore we use the conjugate gradient algorithm to implement the full method reconstruction, both for simulations and experimental data.  

\subsection{Numerical comparison: images}
\label{sec:NumImages}

\begin{figure}[tb!]
\begin{center}
\includegraphics{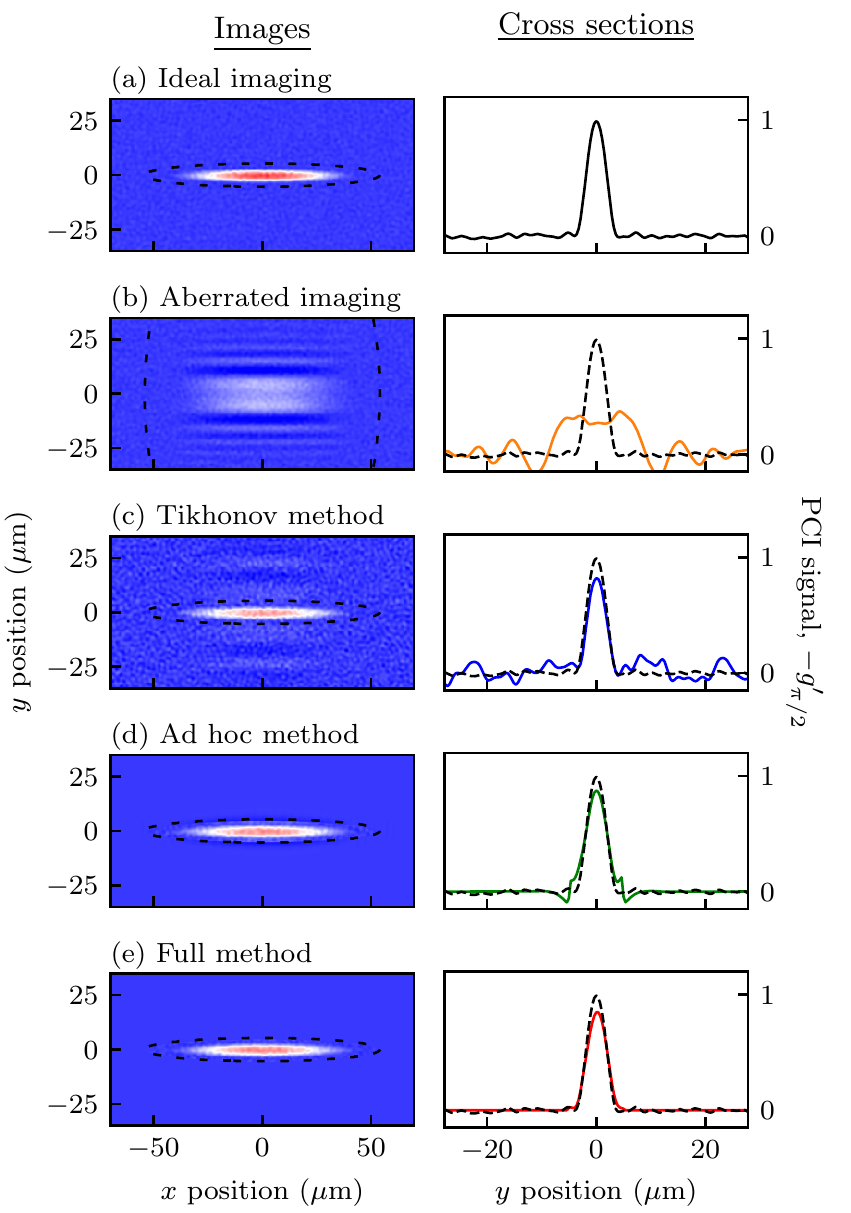}
\end{center}
\caption{Numerically modeled PCI images of $1.1\times10^5$  atoms in a 3D Thomas-Fermi distribution as described in Sec.~\ref{sec:ExpResults}.
In all cases the modeling included photon and atom shot noise and used parameters matching those in our experiments with:
probe detuning $\bar\delta=106$, intensity $\bar I = 2.0$, overall system efficiency of $0.6$ (corresponding to about 400 photo-electrons detected per pixel), and a $20\ \mu{\rm s}$ pulse duration.
Full images are depicted in the left column and vertical cross sections are plotted in the right.
(a) Image following an ideal NA-limited imaging system.
(b) Aberrated image from an imperfect imaging system. 
(c) Reconstruction using the Tikhonov method with $\alpha=0.1$.
(d) Reconstruction using the ad hoc method with $\alpha=0.1$.
(e) Reconstruction using the full method with $\alpha=0.1$.
The dashed curve in each cross section replots the ideal NA-limited case for reference and the dashed black ellipses denote real-space window functions that are relevant both for reconstruction (ad hoc and full method) as well as the computation of the PSD (ideal imaging, imperfect imaging, and Tikhonov method).
}
\label{Fig:Summary}
\end{figure}

In this section we numerically compare the reconstruction methods described above: the conventional spectral ``Tikhonov'' method (Sec.~\ref{subsec:Tikhonov}), the ad hoc method (Sec.~\ref{subsec:adhoc}), and the full method (Sec.~\ref{subsec:full})].

We modeled PCI imaging of an anisotropic BEC with $1.2\times10^5$ atoms and Thomas-Fermi (TF) radii of $R_x = 43.6\ \mu{\rm m}$ and $R_y = 3.5\ \mu{\rm m}$.
In our model, we simulated the imaging system described in Sec.~\ref{sec:Microscope}, with aberration coefficients given in Table~\ref{Table:FitParameters}, and used representative experimental measurement parameters (see Sec.~\ref{sec:ExpResults}); both photon and atom shot noise were included as Poisson random processes.

We use the same overall analysis procedure both for simulated and experimental data:
\begin{enumerate}
\item For each measurement $j$, we obtain three raw images $I^{(j)}_+$, $I^{(j)}_0$, and $I^{(j)}_D$ (for simulated data $I^{(j)}_D$  is not needed).
\item We compute the averaged dark frame $I_D = \langle I^{(j)}_D\rangle$, and remove it from the remaining images: $I^{(j)}_+\rightarrow I^{(j)}_+ - I_D$ and $I^{(j)}_0\rightarrow I^{(j)}_0 - I_D$.
\item To reduce noise and artifacts $I^{(j)}_{\rm PCA}$ is reconstructed using principle component analysis (PCA) techniques~\cite{Li2007,Segal2010} from the full set of $\{I^{(j)}_0\}_j$.
For simulated data there are no imaging artifacts and $I^{(j)}_{\rm PCA}$ is replaced with a modeled shot-noise noise-free probe.
\item We construct the PCI signal $g^{(j)}_{\rm PCI} = 1 - I^{(j)}_+/ I^{(j)}_{\rm PCA}$.
\item Except when otherwise stated, we apply a Fourier window to $g^{(j)}_{\rm PCI}$ describing the known aperture to eliminate photon shot noise present at wavevectors where no signal is present.
\item An image recovery technique of choice (or none at all) is applied to $g^{(j)}_{\rm PCI}$.
\end{enumerate}

The left column of Fig.~\ref{Fig:Summary} depicts modeled PCI data under different conditions.
Panel (a) begins by showing an image from an ideal NA-limited imaging system, while 
(b) introduces aberrations.
Panel (c) shows that conventional Tikhonov reconstruction using $\alpha=0.1$ gives significant added noise and introduces small artifacts parallel to the main reconstitution~\footnote{The regularization parameter $\alpha$ was selected to make the noise and artifacts similar in amplitude.}.
Panels (d) and (e) show reconstructions from the ad hoc and full methods respectively, using an elliptical Tukey window with major and minor axes $(1.25\times R_x, 1.5\times R_y)$ depicted by black ellipses, and Tukey parameter $0.25$.
Both methods appear virtually indistinguishable from the ideal case in (a).
The vertical cross sections plotted in the right column of Fig.~\ref{Fig:Summary} compare the uncorrected data and our three reconstruction methods to the ideal data in more detail; the regularization parameter $\alpha=0.1$ was used in all cases.
The uncorrected data [(b) orange curve] bares virtually no resemblance to the true signal (dashed curve), while the reconstructed signals approximate the true signal with differing degrees of accuracy.
The Tikhonov method [(c) blue curve] accurately recovers the overall shape of the desired distribution, but adds significant noise; increasing $\alpha$ decreases the added noise at the expense of reduced accuracy in the recovered signal.
The ad hoc method [(d) green curve] has greatly reduced noise but introduces artifacts at the edge of the Thomas-Fermi distribution.
Lastly, the full method [(e) red curve] retains the low noise of the ad hoc method while eliminating its artifacts, thereby recovering the true signal with even increased accuracy.
We note that all three of these methods underestimate the PCI signal; this results from the small signal linearization leading to Eqs.~\eqref{eq:field_transfer} and \eqref{Eq:GeneralContrastTransfer}.
In principle this is not needed, but the resulting minimization problem is non-linear and beyond the scope of this paper.

\subsection{Numerical comparison: correlations}

\begin{figure}[tb!]
\begin{center}
\includegraphics{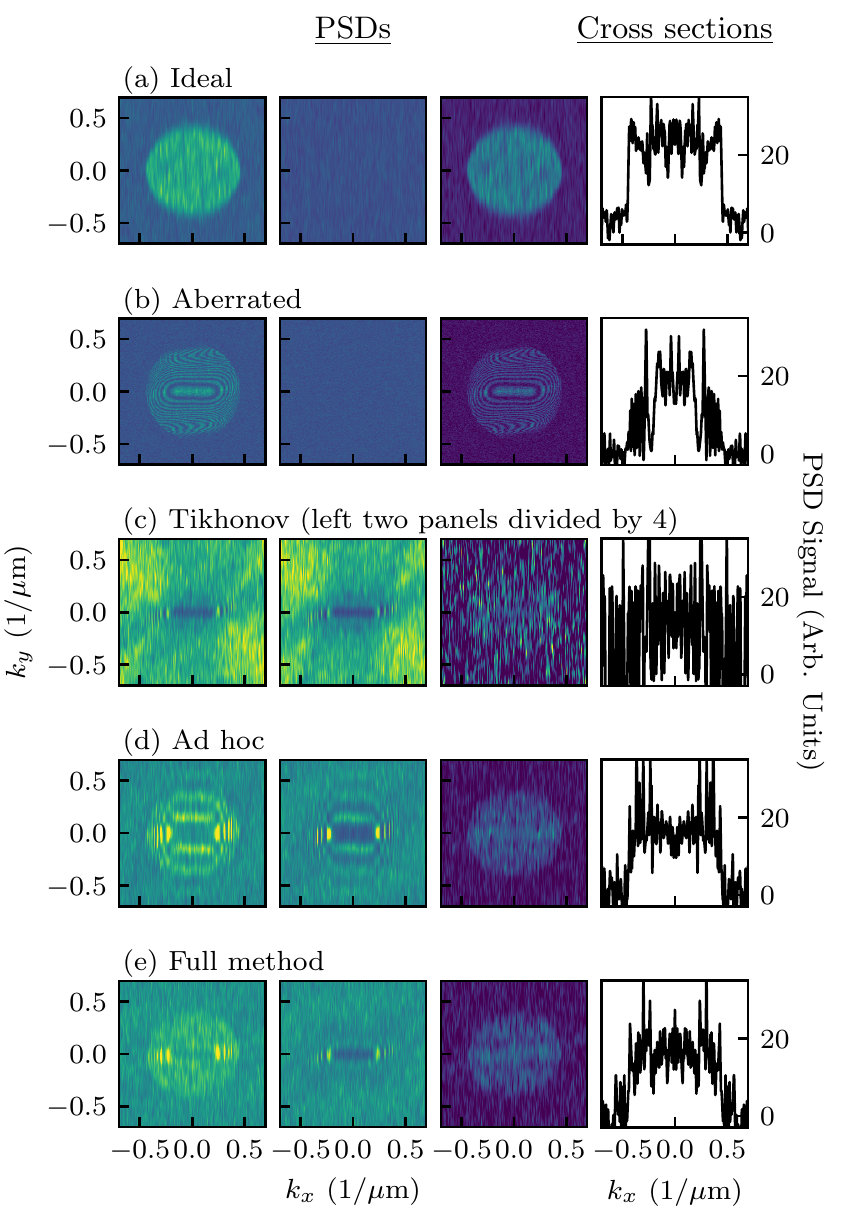}
\end{center}
\caption{Numerically modeled PSDs from images computed as in Fig.~\ref{Fig:Summary} averaged over 100 repetitions.
The left column of figures plot the modeled PSD with atoms present; the next column is the PSD computed with no atoms; and the third column shows their difference.
The right column plots a horizontal cross-section through the difference.
(a) Ideal imaging system.
(b) Aberrated imaging system modeling experimental imperfections.  
These data were processed with a window $10\times$ larger along $\ey$. To compare with the remaining images, this data was simulated with a pulse duration increased by a factor of 10 to an unrealistic $200\ \mu{\rm s}$.
(c)-(e) show the PSD computed following Tikhonov, ad hoc, and full reconstructions, respectively.
}
\label{Fig:SummaryPSD}
\end{figure}

Density-density correlations present in the fluctuations (noise) of cold-atom images can be directly related to the static structure factor~\cite{Hung_2011}.
As established in the previous section, our ad hoc and full methods produce low noise reconstructions; this section takes the next step by analyzing correlations in these reconstructions.
Here we quantify structure in the fluctuations in terms of the PSD given by ${\rm PSD}({\boldsymbol \delta} {\bf d}) \equiv \langle|{\rm FT}({\boldsymbol \delta}{\bf d})|^2\rangle$, where ${\boldsymbol \delta} {\bf d} = {\bf d} - \langle{\bf d}\rangle$ describes the fluctuations observed in a single experiment.
Artifacts in the PSD introduced by imperfect imaging systems can be compensated for~\cite{Hung_2011}; however, previous work did not consider refocusing images.
It is far from clear if refocusing techniques correct correlations, indeed, contrast transfer functions introduce correlations in otherwise uncorrelated noise~\cite{Perry2021}, potentially rendering these methods unsuitable for correlation analyses.
Figure~\ref{Fig:SummaryPSD} illustrates the viability of these refocusing methods via simulations of systems with spatially uncorrelated atom shot noise giving uniform PSDs.

The left panel in Fig.~\ref{Fig:SummaryPSD}(a) plots the PSD resulting from an ideal NA-limited imaging system evidencing signal within a central circle defined by the system's NA, i.e., $|{\bf k}| < k_{\rm NA}$.
Outside this circle, the PSD takes on a non-zero background value from photon shot noise.
The central image plots the PSD when no atoms are present, showing that the photon shot noise signal is constant: as is expected for spatially uncorrelated noise unaffected by the microscope's NA or aberrations.
In these simulations the photon shot noise contribution is minimized by applying the elliptical Tukey window plotted in Fig.~\ref{Fig:Summary}a.  
In this way, photon shot noise from regions with no atoms is eliminated.
Lastly the right image plots the atom-signal alone, obtained by subtracting the PSD with no-atoms (photon shot noise only), from that with atoms (containing signal and photon shot noise).
The final panel plots a horizontal cross-section illustrating the ${\rm SNR}\approx 10$ of the correlations.

Figure~\ref{Fig:SummaryPSD}(b) plots the same quantities computed for our aberrated imaging system showing the appearance of structure in the PSD from aberrations.
These data required a real-space window function $\times 10$ larger along $\ey$ to capture the full diffraction pattern [Fig.~\ref{Fig:Summary}(b)].
To compensate for the added photon shot noise, we increased the imaging pulse duration from $20\ \mu{\rm s}$ to $200\ \mu{\rm s}$.
In practice this imaging time is unrealistically large, so further averaging would be required instead; this makes correlation analyses of highly aberrated PCI images impractical.

Figure~\ref{Fig:SummaryPSD}(c)-(e) addresses the degree to which our regularization methods recover the PSD of the ideal imaging system.
Figure~\ref{Fig:SummaryPSD}(c) shows that the Tikhonov method adds significant structure to the photon shot noise background as well as greatly reducing the SNR of the differenced PSD.
By contrast, (d) and (e) show that the ad hoc and full methods imprint sequentially less structure to the photon shot noise and recover the ideal PSD with increased fidelity.

\begin{figure}[tb!]
\begin{center}
\includegraphics{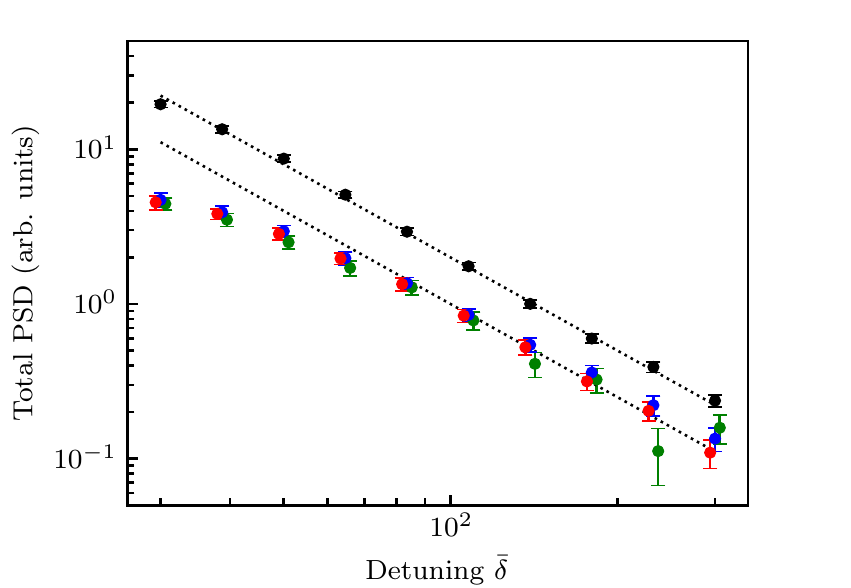}
\end{center}
\caption{Noise analysis of integrated PSD.
Each symbol marks the PSD (computed as described in Fig.~\ref{Fig:SummaryPSD}) integrated within the NA-limited disk.
The black, green, blue and red data respectively plot the results for ideal imaging, the Tikhonov method, the ad hoc method and the full method.  
The dashed lines show the expected $1/\bar\delta^2$ scaling of the PSD signal, with a factor of 2 scale factor between the lines.
}
\label{Fig:IntegratedPSD}
\end{figure}

Figure~\ref{Fig:IntegratedPSD} plots the integrated PSD within the allowed NA window as a function of detuning $\bar\delta$ along with a pair of dashed lines showing the expected $1/\bar\delta^2$ scaling.
The dashed lines differ only by a factor of two, showing that the three reconstruction methods yield a signal about a factor of two below the ideal case, resulting from the actual information last in the process of being aberrated. 
As was anticipated by the individual PSDs, the Tikhonov (green) method exhibits excess noise somewhat in excess of the ad hoc (black) and full (red) methods.
The reduced PSD signal of the reconstructions at small $\bar\delta$ result from the PCI signal $g^\prime_{\pi/2}>1$, invalidating the small-signal approximation used in deriving the CTF.

\section{Ultracold Atom Microscope}
\label{sec:Microscope}

\begin{figure}[t!]
\begin{center}
\includegraphics{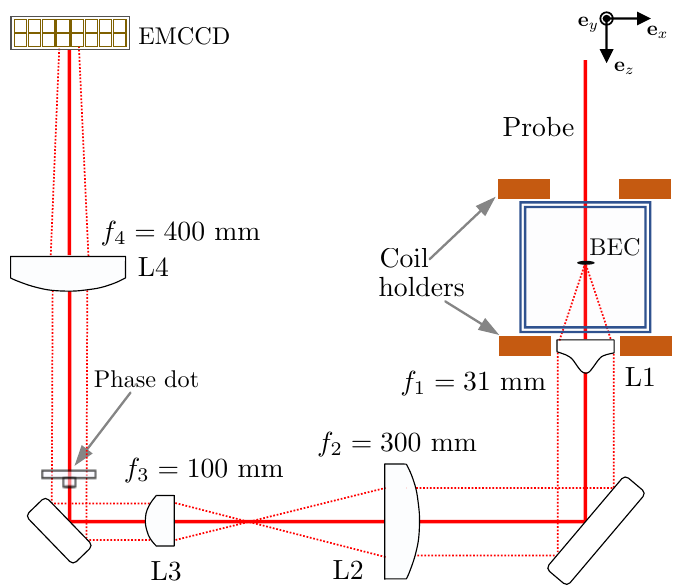}
\end{center}
\caption[Ultracold atom microscope]{Diagram of the ultracold atom microscope.
The cold atoms reside in a vacuum system with a square cross-section glass cell with $\approx50\ {\rm mm}$ sides.
The ${\rm NA} = 0.32$ objective lens maximizes the NA given the constraint of the illustrated coil holders.
Not shown are three additional dichroic mirrors that center the probe beam on the final lens and the EMCCD.
}
\label{Fig:UltracoldMicroscope}
\end{figure}

We imaged BECs at high resolution using an ultracold atom microscope based on a single low cost and NA aspheric lens as the objective lens, shown in Fig.~\ref{Fig:UltracoldMicroscope}.
The optical system consisted of back-to-back Keplerian telescopes with total magnification $\rm{M}=36.3$.
The first stage used an objective lens (L1, with focal length $f_1$, Edmund Optics part number 49-115~\footnote{Certain commercial equipment, instruments, or materials are identified in this paper in order to specify the experimental procedure adequately. Such identification is not intended to imply recommendation or endorsement by the National Institute of Standards and Technology, nor is it intended to imply that the materials or equipment identified are necessarily the best available for the purpose.}) with numerical aperture ${\rm NA}=0.32$. 
The second lens (L2) with $f_2 = 300\ {\rm mm}$ was an achromat with a $50.8\ {\rm mm}$ diameter, selected to minimize vignetting effects. 
The second Keplerian telescope consisted of a pair of lenses (L3 and L4) with focal lengths $f_3 = 100\ {\rm mm}$ and $f_4 = 400~{\rm mm}$. 
The resolution of our microscope, defined by the Rayleigh criterion~\footnote{The Rayleigh criterion is the radius of the first minimum of the NA limited intensity pattern of an imaged point source, i.e., an Airy pattern.}, was diffraction limited with $\approx0.61\lambda/\rm{NA} = 1.5~\mu{\rm m}$ at the imaging wavelength of $\lambda = 780\ {\rm nm}$.
An electron multiplying CCD (EMCCD) with $1024\!\times\!1024$ square pixels (with 13~$\mu$m pixel size) was placed at the image plane located at the focus of L4, where a diffraction limited spot was about $4$ pixels in radius.

Our imaging system included an adjustable mask at the intermediate image plane, allowing us to image elongated atomic ensembles while leaving the majority of the sensor dark.
This enables repeated minimally destructive (ideally quantum back-action limited) measurements of the same ensemble, using the ``fast kinetics mode'' available on some CCD sensors.
All PCI images reported in this paper were taken with the mask fully open, i.e., non-masked and hence the mask is not shown in Fig.~\ref{Fig:UltracoldMicroscope}.

We implemented PCI using a $25.4\ {\rm mm}$ diameter phase plate (manufactured by Lexitek, Inc.) containing a phase dot $37\ \mu{\rm m}$ in radius and $19.5\ \mu{\rm m}$ thick.
This plate was positioned at the Fourier plane of the second Keplerian telescope. 
The nominally Gaussian probe beam, i.e., light which has not been scattered by the atoms was focused by L3 to a $26\ \mu{\rm m}$ $1/e^2$ radius in the phase dot.
By contrast the scattered light was confined to a much larger $\approx 3.8\ {\rm mm}$ radius disk.
As a result virtually all of the unscattered light traveled through the phase dot, while nearly none of the scattered light did.

\section{Experimental Results}
\label{sec:ExpResults}
We imaged highly elongated $^{87}$Rb BECs {\it in-situ} using PCI and PTAI. 
The $N=1.2(2)\times10^5$ atom BECs were created in the $\left|F = 1, m_{F} = 1\right\rangle$ electronic ground state, and were confined in an elongated crossed optical dipole trap (ODT) with frequencies $(\omega_x, \omega_y, \omega_z) = 2\pi \times \left[ 12.2(1), 153.2(3), 175.4(5) \right]$ Hz.
We obtained $N$ from the {\it in-situ} longitudinal TF radius $R_x = 43.6(9)~\mu{\rm m}$~\cite{Dalfovo1999}, resulting in $R_y = 3.5(1)~\mu{\rm m}$ and $R_z = 3.0(1)~\mu{\rm m}$.  
In addition we applied the Castin-Dum scaling theory~\cite{CastinDum96} to separately measured time-of-flight (TOF) images, and found $N=1.9(3)\times10^5$, which would imply an $R_x = 48(2)~\mu{\rm m}$ that is inconsistent with our {\it in-situ} observations.

Our probe laser couples the ground $\ket{F=2,m_F=2}$ state to the excited $\ket{F^\prime=3,m_F^\prime=3}$ state.
As a result, we transferred the atoms from $\ket{F=1,m_F=1}$ to $\ket{F=2,m_F=2}$ using a $68\ \mu{\rm s}$ resonant microwave pulse prior to PCI imaging.
For PTAI we used a weaker microwave pulse to transfer $\approx10~\%$ of the population to $\ket{F=2,m_F=2}$.
In both cases, the imaging pulse was $20\ \mu{\rm s}$ in duration and had intensity $I/I_{\rm{sat}}\approx 2$, where $I_{\rm{sat}} \approx 1.67~\rm{mW/cm^2}$.

Our near-resonance ``absorption imaging'' measurements were altered by the presence of a phase dot in our microscope.
The OD in this case is given by
\begin{equation}
{\rm OD}_{\rm AI}({\bf r}_\perp) = \frac{1}{2\bar \delta}g^{\prime}_{\theta=\pi/2}({\bf r}_\perp),
\label{eqn:OD_AI_phaseDot}
\end{equation}
where we evaluated Eq.~\eqref{eqn:g_General} assuming both ${\rm OD} \ll 1$ and  $\bar\delta \ll 1$. 
Interestingly this is the same expression as for PCI given in Eq.~\eqref{eqn:OD_PCI_piOver2}, although the resulting signal is from absorption not phase shift.

The remainder of this section proceeds as follows.
First we describe our experimental protocol extending Eq.~\eqref{Eq:Pupil_PSD} for characterizing the microscope's Fourier pupil function using PSDs obtained from near resonant PTAI images.
We then contrast high-resolution PCI images of our BEC reconstructed using the standard Tikhonov method with those from our full method.
We conclude by applying our full method to {\it in-situ} imaging of the thermal to BEC phase transition, which is difficult to resolve in our aberrated raw data.

\subsection{Fourier pupil function measurements}
\label{sec:ExpPupilFunc}

We experimentally characterized the Fourier pupil function of our ultracold atom microscope utilizing density-density correlations and the BEC's TF distribution.
As discussed in Sec.~\ref{Section:Aberrations}, PSDs provide information about aberrations present in imaging systems.  
We extracted density correlations in the fluctuations of cold-atom images and obtained experimental PSD similar to the numerical model shown in Fig.~\ref{Fig:SummaryPSD} (b). 
The PSD contains no information about the anti-symmetric phase $\beta_-$ contributions to the pupil function, and instead we used the difference between the reconstruction and the expected TF distribution to constrain $\beta_-$.

Our strategy for measuring the Fourier pupil function via PSDs combines two critical elements to deliver increased precision.
First, we obtained the PSD from {\it in-situ} PTAI images.
As described in Sec.~\ref{Section:Absorption Imaging}, PTAI introduces uncorrelated atom shot noise to deeply degenerate BECs; the observed PSDs then carry the imprint of our microscope's aberrations upon a featureless background.
Second, we deliberately defocused our microscope by translating L4, the lens immediately preceding the EMCCD (see Fig.~\ref{Fig:UltracoldMicroscope}), away from the established focal position at $\delta z_{\rm{L}4} = 0\ \rm{cm}$.
Changing the focus by a small distance $z$ adds a quadratic phase shift $z k_\perp^2 / 2 k_{0}$ to the pupil function as introduced in Sec.~\ref{Section:Aberrations}. 
Then PSD measurements taken at different image planes differ only in their $k_\perp^2$ terms.
Consequently by performing a joint fit to a family of such PSDs we quantified the imaging system's even-order aberrations with increased precision.

\subsubsection{Correlations fit function}
Following the aberration model discussed in Sec.~\ref{Section:Aberrations}, we employed a fit function that accounts for optical aberrations as well as unwanted ``surface effects'' (including reflections, along with losses within the optical elements), and aperture limits.
The attenuation parameter $\gamma_+({\bf k}_\perp)= \gamma_+^{\rm S}({\bf k}_\perp) + \gamma_+^{\rm A}({\bf k}_\perp)$ describes the exit pupil apodization, where we have introduced surface and aperture contributions $\gamma_+^{\rm S}$ and $\gamma_+^{\rm A}$.
Because our imaging system is well aligned on the optical axis, we assume $\gamma_-({\bf k}_\perp)$ has no surface components, i.e., $\gamma_-({\bf k}_\perp) \equiv \gamma_-^{\rm A}({\bf k}_\perp)$. 
These variables allow us to re-express Eq.~\eqref{Eq:Pupil_PSD} as
\begin{align}
	\langle |\delta {\rm OD}({\bf k}_\perp) |^2 \rangle \propto &\  e^{-2\gamma_+^{\rm S}({\bf k}_\perp)}\bigg\{\frac{1}{2}[ A^2({\bf k}_\perp) +  A^2(- {\bf k}_\perp)]\nonumber\\
	&+ h_{\rm dof}({\bf k}_\perp)
	[A({\bf k}_\perp) A(- {\bf k}_\perp)]	\nonumber\\
	&\times \cos[2\beta_+({\bf k}_\perp) + 2(\varphi-\theta)]\bigg\} 
	\label{Eq:PupilPSD_fitFunc},
\end{align}
where $A({\bf k}_\perp) \equiv e^{-\gamma^{\rm A}({\bf k}_\perp)}$.
We interpret $A({\bf k}_\perp)$ as a window describing the aperture~\footnote{We model $A({\bf k}_\perp)$ as a boxcar window function that takes on values of either $1$ (inside) or $0$ (outside) so $A^2=A$.}.
We empirically determined $A({\bf k}_\perp)$ based on prominent structures in the measured PSD that result from the known experimental geometry of our apparatus.
The details of this procedure are given in the following section. 

Second, we characterize the phase shift of the Fourier pupil function using the polynomial representation
\begin{equation}
	\beta({\bf k}_\perp) = \sum_{m,n} c_{mn} \left(\frac{k_x}{k_0}\right)^m  \left(\frac{k_y}{k_0}\right)^n.
	\label{Eqn:PupilPhasePoly}
\end{equation} 
The PSD depends on $\beta_+$, thus our fit function contains only symmetric terms, i.e., those with even $m+n$.
We thereby model even-order aberrations such as astigmatism, defocus and spherical aberrations manifested in our microscope.
Because our experimental aperture (described below) is not circular, the conventional Zernike basis has no particular meaning.
While it would in principle be possible to construct an orthogonal polynomial basis for our aperture, we adopt a simple order-by-order polynomial expansion.

We performed a global fit of all $\langle |\delta {\rm OD}({\bf k}_\perp) |^2 \rangle$ measurements discussed in the next section to Equ.~\eqref{Eq:PupilPSD_fitFunc} using the following criteria.
The degree of defocus is quantified by $c_{20}$ and $c_{02}$; these change by the same amount as $\delta z_{{\rm L}4}$ changes, but the remaining $c_{mn}$ are fixed.
As a result, the global fit designates $c_{20}$ and $c_{02}$ as unshared parameters (constrained to change by the same amount as $\delta z_{{\rm L}4}$ changes) while the remaining $c_{mn}$ are held constant across the data sets.
We performed two supplementary measurements at $\delta z_{{\rm L}4} = 0$ by changing the detuning to $\bar \delta \approx \pm 0.5$.
Equation~\eqref{Eq:PupilPSD_fitFunc} shows that $c_{00}=(\varphi-\theta)$ results from detuning and the PCI phase shift.
The fits to these supplementary measurements share all their parameters with the $\bar\delta=0$ dataset except $c_{00}$.

The surface term in Eq.~\eqref{Eq:PupilPSD_fitFunc} is independent of $\delta z_{{\rm L}4}$ and $\bar\delta$, and we represent it as a Gaussian using $\gamma_+^{\rm S}({\bf k}_\perp) = {g_{\rm S}}^2 [(k_x/k_0)^2 + (k_y/k_0)^2]$, where ${g_{\rm S}}$ is a shared fit parameter in our aberrations model.
Finally, following Eq.~\eqref{eqn:f_eff_2}, the DoF term $h_{\rm dof}$ is parameterized by the shared fit coefficient $c_{\rm dof} \equiv w_z k_0/4$, which depends on the thickness of the cloud $w_z$ in the imaging direction.
We include this effect in our fits, but the resulting $w_z\approx 18 ~\mu\rm{m}$ is far from $R_z$, implying that oscillatory structure is lost for reasons other than the DoF effect.
For example the field of view discussion in Appendix~\ref{app:grid} implies such an effect.

\subsubsection{Density correlation measurements}

\begin{figure}[tb!]
	\begin{center}
		\includegraphics{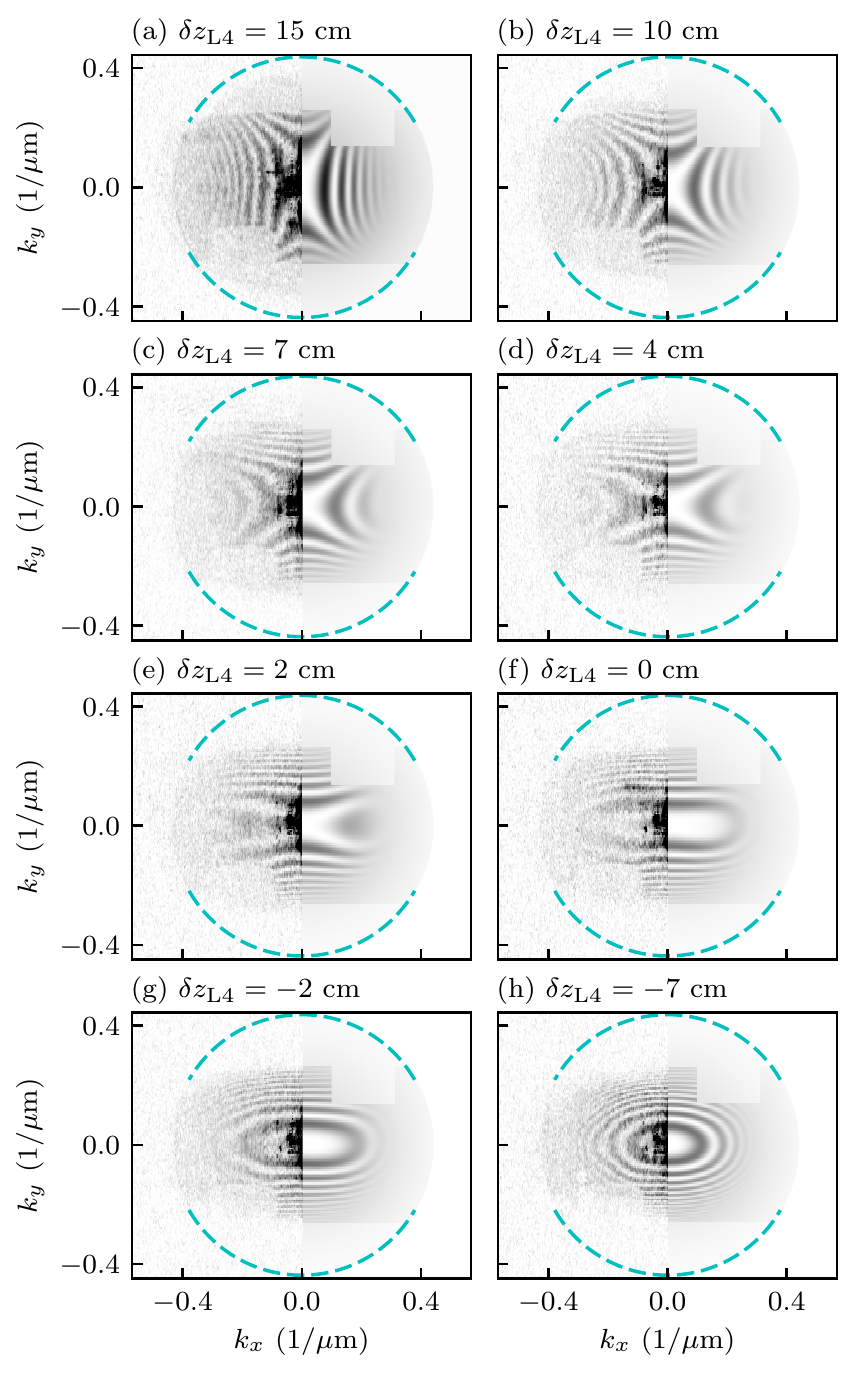}
	\end{center}
	\caption{PSDs $\langle |\delta {\rm OD}_{{\bf k}_\perp}|^2 \rangle$ at differing degrees of defocus including experimental data (left, $k_y<0$) and fits (right, $k_y>0$). 
	These data represent our full set of $\delta z_{\rm{L}4}$ values, and each measurement was averaged over 100 to 200 images.
	The dashed arcs on the top and bottom plot the NA limit $|{\bf k}_\perp| = k_{\rm NA}$ expected for our objective lens.
	The values for the model parameters in the fits are given in Table~\ref{Table:FitParameters} and Fig.~\ref{Fig:xyDefocus_parameters}.
	}
	\label{Fig:ExpdPSD_L4_adjusted}
\end{figure}

\begin{figure}[t!]
\begin{center}
\includegraphics{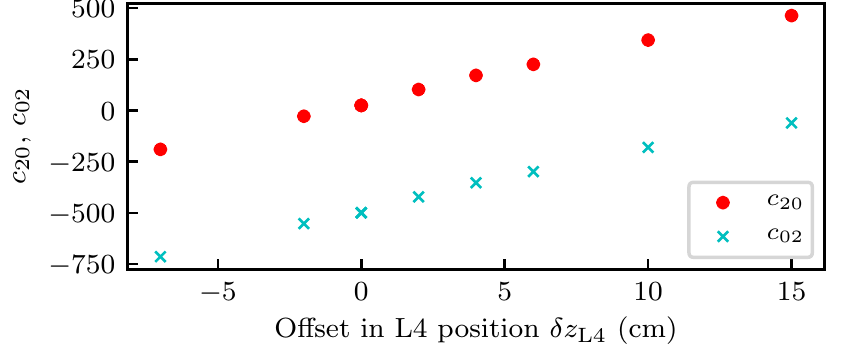}
\end{center}
\caption{Best-fit values for $c_{20}$ and $c_{02}$, extracted from fits to PSD measurements, plotted as a function of $\delta z_{\rm{L}4}$. 
At the nominal $\ex$ focal position $\delta z_{\rm{L}4} = 0$ the coefficients are $c_{20} = 24.9(5)$ and $c_{02} = -500(1)$. 
}
\label{Fig:xyDefocus_parameters}
\end{figure}

Figure~\ref{Fig:ExpdPSD_L4_adjusted} shows PSDs measured from {\it in-situ} PTAI images of BECs taken at a range of image planes (left half of plots, i.e., $k_y<0$) along with global fit to the aberrations model in Eq.~\eqref{Eq:PupilPSD_fitFunc} (right half of plots, i.e., $k_y>0$).
The best-fit values for the shared parameters are reported in Table~\ref{Table:FitParameters}.
The defocus parameters $c_{20}$ and $c_{02}$ are shown in Fig.~\ref{Fig:xyDefocus_parameters} as a function of $\delta z_{\rm{L}4}$~\footnote{In total, the global fit to 11 data sets had 40 fit parameters - including overall amplitude and background offset terms for each data set that are not reported.}.

We determined the aperture term $A({\bf k}_\perp)$ for the fit via the following procedure. 
The overall numerical aperture of the main objective lens limits the maximum accepted wavevector to $k_{\rm{NA}}$ (dashed cyan arcs in Fig.~\ref{Fig:ExpdPSD_L4_adjusted}) and thereby $\langle|\delta {\rm OD}({\bf k}_\perp)|^2\rangle\rightarrow 0$ for $|\mathbf{k}_\perp| > k_{\rm NA}$. 
We observe a  non-zero background outside the NA circle, as expected from photon shot noise.
Our PSD measurements exhibit additional structures, and we focus on the pair at positive $k_y$ giving additional limits to the effective vertical NA (because the PSD derives from the Fourier transform of a real valued quantity, the structures at $k_y < 0$ replicate those at $k_y>0$).
First, the horizontal cutoff at  $k_y \approx 0.26~\mu\rm{m}^{-1}$ results from an in-vacuum ``atom-chip'' in our apparatus that intercepts wave-vectors at large $k_y$.
A second rectangle carved into the aperture results from screw heads extending down from the atom-chip holder.
Extending the dashed cyan curves in Figure~\ref{Fig:ExpdPSD_L4_adjusted} shows that the expected NA limited disk is present for small $|k_y|$ where the atom-chip NA limitations are not present.
In our fit $A({\bf k}_\perp)$ is modeled as a window function that combines the NA disk of the objective lens with the two additional vertical aperture limits resulting from the atom-chip assembly at positive ${\bf k}_\perp$. 
While all the data in Fig.~\ref{Fig:ExpdPSD_L4_adjusted}(a) have NA limits from the atom-chip assembly, the effects are most visible in (a) which is nearly in focus along $\ey$.
We therefore determined the aperture window function from the PSD signal in Fig.~\ref{Fig:ExpdPSD_L4_adjusted}(a). 

\begin{table*}[tb!]
\begin{tabular}{ c c c c c c c c c c c} 
\hline\hline
 Parameter & $c_{00}$   &  $c_{11}$    &    $c_{03}$   &  $c_{31}$   &   $c_{13}$  & $c_{22}$   &  $c_{40}$    &    $c_{04}$   &  $\rm{c}_{\rm dof}$   &   ${g_{\rm S}}$ \\
\hline
 Value    & 1.590(4) &  -43.3(4)  & -0.52(2)\e{3} & 1.82(2)\e{3}  & 2.71(3)\e{3} &
  -0.26(4)\e{3}  &  -1.85(2)\e{3}  & -3.35(2)\e{3} & 35.9(1)  & 3.033(2)  \\
\hline
\hline
\end{tabular}
	\caption{Best-fit parameter values.
    The shared global parameters ${g_{\rm S}}$, $\rm{c}_{\rm dof}$ and $c_{mn}$ (for $n,m\geq0$ and even $n+m$) result from our PSD fits.
	We additionally include $c_{03}$ derived from coordinate space TF fits.    
	All coefficients are dimensionless. 
	}
	\label{Table:FitParameters}
\end{table*}

Equation~\eqref{Eq:PupilPSD_fitFunc} describes two key features of the aperture limits that stem from the atom-chip assembly. 
First, although only up-going scattered light is blocked by then atom-chip assembly, we observe the atom-chip NA limit for both positive and negative $k_y$.
In Eq.~\eqref{Eq:PupilPSD_fitFunc}, the first term in curly brackets is a symmetrized aperture that terminates the non-oscillatory contribution to the PSD.
This eliminates correlations outside $k_{\rm NA}$ disk in the experimental data.
In the second term, the product $A({\bf k}_\perp) A(-{\bf k}_\perp)$ predicts that the oscillatory structure given by $\cos(\cdots)$ terminates at the aperture boundaries. 
This is observed at the aperture limit from the atom chip assembly as well as the NA limit near $k_y = 0$.

The magnification of our microscope changes as a function of $\delta z_{{\rm L}4}$; at $\delta z_{\rm{L}4} = 0\ \rm{cm}$ the resolution is given by the design magnification $M=36.3$. 
We empirically identified the magnification at each $\delta z_{\rm{L}4} \neq 0$ by aligning the observed and expected NA circles.
All of our data is presented including these calibrated magnifications.

Similar to the simulated data analysis, photon shot noise was subtracted from the experimental PSD to isolate the atom shot noise.
We begin by masking out the signal inside the expected NA circle, where atom shot noise is dominant.
We then average the masked data along $k_y$ and subtract it from the signal (eliminating structured noise along $k_x$).
Next we repeat the same subtraction procedure by averaging along $k_x$ (eliminating structured noise along $k_y$).

As demonstrated in Fig.~\ref{Fig:ExpdPSD_L4_adjusted}, our aberrations model, using coefficients from our global fit, accurately characterizes our microscope and consistently describes the observed aspects in all PSD measurements. 
The imaging system is astigmatic: Figs.~\ref{Fig:ExpdPSD_L4_adjusted}(a) and (f) show data nearly focused along $\ey$ and $\ex$ respectively, where the $c_{02}$ and $c_{20}$ coefficients approach zero in Fig~\ref{Fig:xyDefocus_parameters}. 
Hence panel (a) has relatively little oscillatory structure along $\ey$, but significant structure along $\ex$; this pattern reverses progressively from (a) to (f) as $\delta z_{{\rm L}4}$ decreases.
The remaining data (g) and (h) show increasing oscillatory structure in both directions as $\delta z_{\rm{L}4}$ becomes more negative.

Our global fit provides a measure of our phase dot's phase shift $\theta$ using $c_{00}$ obtained for $\bar\delta=0$ along with those measured at $\bar\delta=\pm1/2$.
The best-fit values of $c_{00}$ are $\left\{1.510(6), 1.590(4), 1.812(5) \right\}$ for detunings  $\left\{0.5, 0, -0.5\right\}$ respectively.
The fit function linearizes Eq.~\eqref{eqn:g_General} around a non-zero optical depth, avoiding the $1/\bar\delta$ divergence in the small OD expression.
Combining these data gives $\theta = -1.6(1)\ {\rm rad}$, which is in good agreement with the design value of $|\pi/2|$ further demonstrating the accuracy of our measurement protocol and aberrations model. 
We also note that the detunings are offset by $\bar \delta = -0.03(6)$. 

\subsubsection{Determining anti-symmetric pupil phase contributions}

Imaging aberrations determined from PSD measurements yield all components of the pupil function except the anti-symmetric $\beta_-$ described by the odd order $c_{mn}$ parameters.
In our data, images reconstructed with $\beta_-=0$ have asymmetric dips above and below the central density peak.
We determined $c_{03}$ term by minimizing the difference between reconstructed images and the expected TF distribution.
We omitted the first-order terms as they describe real-space translations.
Because our BEC's density distribution is highly elongated along $\ex$, its spectral distribution contains only small $k_x$ components.
As a result, only coefficients $c_{0m}$ significantly alter the overall density distribution. 
We then fit reconstructed images to the 2D TF distribution 
\begin{align}
\rho(y) &= \rho_0 \left[1-\left(\frac{x-x_0}{R_x}\right)^2 -\left(\frac{y-y_0}{R_y}\right)^2 \right]^{3/2}
\end{align}
with $c_{03}$ (the lowest order remaining contributor to $\beta_-$) included as a fit parameter.
The best-fit value for $c_{03}$ is given in Table~\ref{Table:FitParameters}.

\subsubsection{Final pupil model}

\begin{figure}[t!]
\begin{center}
\includegraphics{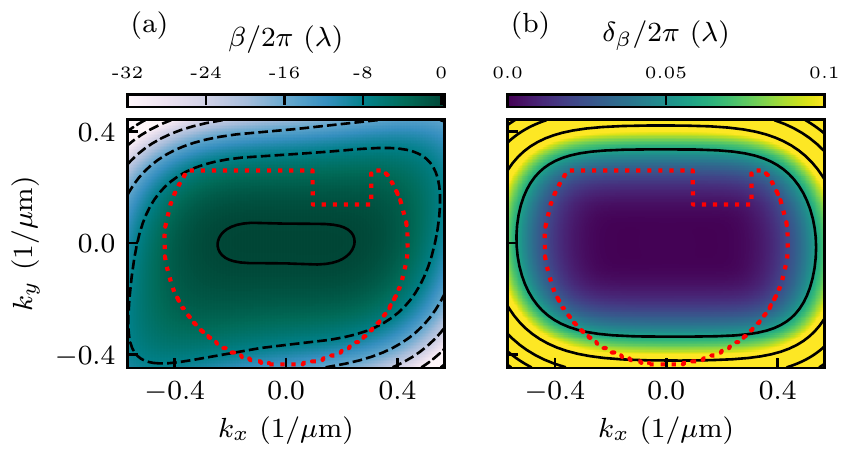}
\end{center}
\caption{Final pupil wavefront model at $\delta z_{{\rm L}4} = 0\ {\rm cm}$. 
(a) Pupil phase wavefront mean and (b) standard deviation.
The red dashed curve outlines the complete aperture limit due to the ultra-high vacuum apparatus geometry.
Contour lines (black) are spaced approximately every $8\lambda$ in (a) and $\lambda/20$ in (b).
}
\label{Fig:PupilWavefront_BestFocus}
\end{figure}

Figure~\ref{Fig:PupilWavefront_BestFocus}(a) presents our final model for the pupil phase wavefront $\beta$ evaluated at $\delta z_{\rm{L}4} = 0\ \rm{cm}$. 
Figure~\ref{Fig:PupilWavefront_BestFocus}(b) plots the uncertainty $\delta_{\beta}({\bf k}_\perp)$ computed from our fits' combined covariance matrix (with a total of 41 parameters including shared parameters) assuming a multivariate normal distribution of parameters. 

For a complete model of $\beta$, this would imply an rms wavefront error $0.03 \lambda$ associated with reconstructed images.
In our demonstrated fourth-order model, we were unable to model the $c_{12}$, $c_{21}$ and $c_{30}$ coefficients, which contribute unknown wavefront errors, implying that $0.03 \lambda$ is a lower bound for the rms wavefront error of our reconstructions. 

\subsection{Digitally enhanced non-destructive imaging with far-detuned PCI}
\label{sec:ExpRegularized}

\begin{figure}[t!]
\begin{center}
\includegraphics{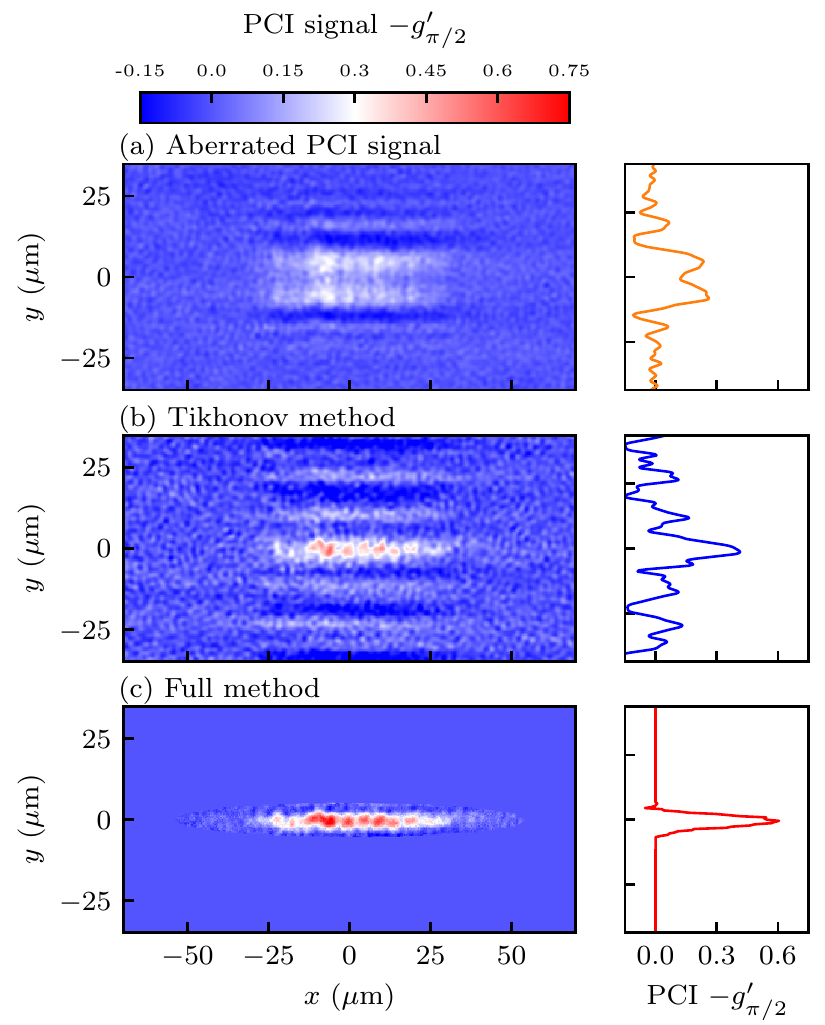}
\end{center}
\caption{Far detuned {\it in-situ} PCI images of BECs showing raw and reconstructed signals. 
(a) Raw PCI signal at probe detuning $\bar\delta \approx 106$.
(b) Reconstructed PCI signal with the Tikhonov approach with $\alpha = 0.1$. 
(c) Reconstructed PCI signal using the full method.
For each case a vertical cross section is shown on the right. 
}
\label{Fig:RegularizedPCI}
\end{figure}

\begin{figure*}[t]
\begin{center}
\includegraphics{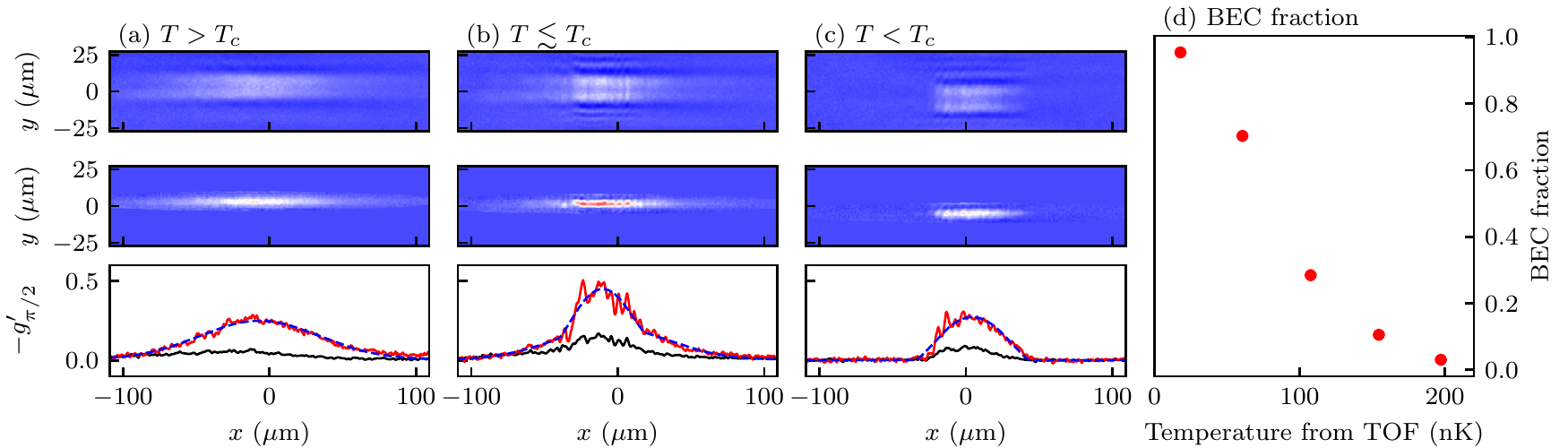}
\end{center}
\caption{Direct {\it in-situ} observation of BEC phase transition.
Top and middle: raw and refocused images.
Bottom: horizontal cross sections for the raw (black) and reconstructed (red) data, and the fits (purple dashed) to 1D Gaussian, bimodal and TF distributions respectively for each regime. 
Each raw image was acquired at probe detuning $\bar\delta \approx 106$ and averaged over 20 images. 
(a) Thermal cloud at $T>T_c$. 
(b) Partially condensed system at $T\lesssim T_c$, with both thermal and condensate components visible.
(c) Pure condensate at $T\ll T_c$.
(d) Condensate fraction measured from {\it in-situ} images plotted as a function of temperature obtained separately from TOF data.
}
\label{Fig:inSitu_BEC_obs}
\end{figure*}

With the aberrations of our ultracold atom microscope quantified, we proceed to aberration compensation of images of BECs taken {\it in-situ} with far-detuned PCI.
Figure~\ref{Fig:RegularizedPCI}(a) presents the raw aberrated image, while (b) and (c) compare reconstructions using the Tikhonov (with $\alpha = 0.1$) and full methods. 
The observed background noise in the aberrated image (a) is consistent with that predicted by our numerical model [Fig.~\ref{Fig:Summary}(b)].
The full method used a 2D elliptical Tukey window function with semi-major and semi-minor axes $(1.25\times R_x, 1.5\times R_y)$, and with Tukey parameter $0.25$; $R_x$ and $R_y$ are the TF radii determined {\it in-situ}.
The Tikhonov reconstruction contains multiple artifacts and added noise, and as discussed in Sec.~\ref{sec:NumImages}, $\alpha$ in Eq.~\eqref{eq:Tikhonov_2} presents a trade-off: noise is reduced, but the accuracy of the reconstruction is sacrificed.
On the other hand, the full method reduces both noise and spectral artifacts while recovering the TF distribution with increased accuracy.

Our reconstruction does not include the experimentally determined aperture $A({\bf k}_\perp)$ in the contrast transfer function $h({\bf k}_\perp)$.
Both with experimental and simulated data, including the rectangular structure from the atom chip assembly led to significant artifacts in the ad hoc reconstruction and somewhat degraded the performance of the full method. 

\subsubsection{{\it In-situ} Observation of BEC Phase Transition}
\label{sec:MethodApplication}

Here we demonstrate an application of increased accuracy of the full regularization method by directly and non-destructively observing condensate formation in a crossed ODT using far-detuned PCI.
Figure~\ref{Fig:inSitu_BEC_obs} reveals the BEC phase transition in the refocused images (middle row) as we decrease the ODT depth, cooling to lower temperatures from above the critical temperature $T>T_c$ in (a), to just below $T\lesssim T_c$ in (b) and to well below $T\ll T_c$ in (c). 
We independently imaged the cold cloud in time-of-flight using AI to calibrate the temperature. 

In raw aberrated images (Fig.~\ref{Fig:inSitu_BEC_obs} top row) only very qualitative features of the density distribution are visible, stymieing quantitative analysis.
The bottom row of Fig.~\ref{Fig:inSitu_BEC_obs} compares the horizontal cross sections of raw images (black curves) and refocused images (red curves) and the fits (purple dashed curves) to the expected density profile for each case. 
We observe that the refocused data are generally in good agreement with the expected thermal plus TF distribution.  
However, in both cases with $T<T_c$, we observe oscillatory structure in the density around $x\approx -20\ \mu{\rm m}$, potentially indicating a previously undetected fringe on our ODT laser beam.
Lastly, Fig.~\ref{Fig:inSitu_BEC_obs}(d) shows the condensate fraction obtained from our {\it in-situ} non-destructively measured yet aberrated images, illustrating the effectiveness of our reconstruction method to yield images suitable for quantitative analysis.

\section{Conclusion and outlook}
\label{sec:Conc}

In this paper we presented a versatile high-resolution ultracold atom microscope composed of two main components: 1) an economical and practical imaging system based on high NA of-the-shelf optics; and 2) a novel, high-fidelity digital aberration removal technique that is compatible with a wide range of imaging techniques.
The combination of these two elements yields an ultracold atom microscope that can be easily integrated to existing cold-atom apparati, this is in contrast with quantum gas microscopes, which necessitate costly and custom designed optics.
Imaging artifacts resulting from the geometrical constraints of an existing vacuum system or imperfections in the optical elements are mitigated using our digital aberration removal technique.
As such our high-resolution ultracold atom microscope is adaptable, simple and effective. 
Furthermore, our reconstruction algorithms are not limited to cold-atom experiments and can be applied in any case where the real and imaginary parts of the susceptibility are proportional to the quantity of interest.

Our full method completely solves the minimization problem at the price of a numerically costly iterative algorithm.
We also showed that a simple ad hoc approximation leads to a method with only slightly degraded performance, suitable for real-time use in a lab setting.

All of our current implementations approximate the true relationship between the detected signal and the ideal recovered signal with a linear transformation that is valid only for small signals.
This leads to the visible underestimation of the true density in the simulated reconstructions which have peak signal $g\approx1$.
Although it is doubtful that algebraic progress beyond Eq.~\eqref{eq:objective} can be made for the true non-linear transformation, we expect that non-linear numerical methods would be able to find the recovered signal without the small $g$ approximation.
This would extend this method to be applicable to the full range of available data.

\begin{acknowledgments}
We benefited greatly from discussions with L.~Walker, R.~Lena, S.Flannigan, A.~Daley, and  W. D. Phillips.
This work was partially supported by NIST, and the NSF through the Physics Frontier Center at the JQI. 
\end{acknowledgments}


\appendix

\section{Series expansion}\label{app:series}

The numerator of Eq.~\eqref{eq:MinErrorZero} can be evaluated using Fourier methods, but because an inverse is required, the denominator is difficult to evaluate.
Reference~\cite{Sprent1965} (see page 186) showed that nearly diagonal matrices have a compact series expansion that in the present case allows for (somewhat) efficient evaluation.
One expression for the inverse is 
\begin{align*}
\left(1 + {\bf J}^\dagger {\bf J} \right)^{-1} &\approx {\bf D}^{-1} - {\bf D}^{-1} {\bf N} {\bf D}^{-1} \\
&+ {\bf D}^{-1} {\bf N} {\bf D}^{-1} {\bf N} {\bf D}^{-1} - \cdots
\end{align*}
where ${\bf D}$ denotes any matrix of diagonal elements, where ${\bf 1} + {\bf J}^\dagger {\bf J} = {\bf D} + {\bf N}$.
This whole expansion may be computed in a straightforward manner, and while this method converges, it does so slowly.
The art in this method is in the selection of ${\bf D}$ to give the most rapid convergence.
We found superior performance using the conjugate gradient method described in the main text.

\section{Grid size and padding}\label{app:grid}

Our method acquires additional considerations when the field of view is limited, i.e., when a significant fraction of the aberrated diffraction pattern is outside the observed field of view.
Here we consider this case by analyzing Eq.~\eqref{Eq:GeneralContrastTransfer}.

In the general vicinity of some ${\bf k}_0$ the phase shift may be Taylor expanded as $\beta({{\bf k}_0 + \delta{\bf k}}) \approx \beta({{\bf k}_0}) + \delta{\bf k}\cdot\nabla_{\bf k} \beta({\bf k}$); thus both terms in $h({\bf k})$ are approximated by displacement operators, with $\delta {\bf x} = \pm \nabla_{\bf k} \beta({\bf k})$, for Fourier components centered at ${\bf k}_0$. Our data consists of images with extent $L$; assuming the object is centered on the image, this implies that for $\delta {\bf x} > L/2$ the information near ${\bf k}_0$ will not have been detected.

Our algorithm uses standard Fourier methods with periodic boundary conditions, in which case these components will wrap-around: a non-physical behavior.
To avoid this, we require $|\partial_{k_x,k_y} \beta({\bf k})| < L/2$; when we  discretize onto a momentum lattice with spacing $2\pi/L$ this implies
\begin{align}
    |\beta({k_0 + 2\pi/L}) - \beta({k_0})| &< \pi.\label{eq:app:condition}
\end{align}
This is to say any phase change in a single momentum-space pixel that is larger than $\pi$ will give signal outside the field of view and should not be included.
In other language, this is the Nyquist limit associated with this signal.

In our implementation we resolve this two ways: 
(1) We cap the gradient of the phase shift $\nabla_{\bf k}\beta({\bf k})$ as it approaches the Nyquist threshold.
(2) If this is insufficient (for example, if the Nyquist limit is violated inside the imaging aperture), we pad the measurement ${\bf m}$ such that Eq.~\eqref{eq:app:condition} is satisfied, and set the inverse uncertainties $\bar\sigma^{-1}$ to zero at these points, thereby assigning them zero weight in the objective function. 

\bibliography{BayesianRefocus}

\end{document}